\UseRawInputEncoding

\documentclass[10pt, prb, aps, twocolumn, showpacs, citeautoscript, floatfix, reprint, amsmath, amssymb, notitlepage, superscriptaddress]{revtex4-1}

\usepackage{booktabs}
\usepackage[utf8]{inputenc}
\usepackage[T1]{fontenc}
\usepackage[english]{babel}
\usepackage{graphicx}
\usepackage{rotating}
\usepackage{mathtools}
\usepackage{amsfonts}
\usepackage{dcolumn} % align table columns on decimal point
\usepackage{bm} % bold math
\usepackage{xcolor}
\usepackage{latexsym}
\usepackage{wasysym}
\usepackage[colorlinks, citecolor=blue, linkcolor=blue, urlcolor=blue]{hyperref}
\usepackage[nomain, acronym, shortcuts]{glossaries}
\usepackage{tikz}
\usepackage{environ}
\usepackage{csquotes}
\usepackage[normalem]{ulem}
\usepackage{amsmath}

\glsdisablehyper
\usetikzlibrary{external}
\bibpunct{[}{]}{,}{n}{}{}

\begin{document}

\title{Magnetoelectric torque and edge currents caused by spin-orbit coupling}

%\title{Room temperature persistent current induced by magnetization and Rashba spin-orbit coupling or interorbital hopping}

%\title{Room temperature persistent current induced by Rashba spin-orbit coupling and magnetization}

%\title{Persistent charge current in magnetized 2D spin-orbit systems}

%\title{Persistent charge current induced by the interplay of magnetization and spin-orbit coupling}

%\title{Persistent charge current induced by inverse Edelstein effect}

\author{Wei Chen}

\affiliation{Department of Physics, PUC-Rio, 22451-900 Rio de Janeiro, Brazil}

\author{Manfred Sigrist} 

\affiliation{Institute for Theoretical Physics, ETH Zurich, 8093 Zurich, Switzerland}

\date{\rm\today}

\begin{abstract}

Using a tight-biding model, we elaborate that {  the previously discorvered out-of-plane polarized helical edge spin current caused by Rashba spin-orbit coupling can be attributed to the fact that} in a strip geometry, a positive momentum eigenstate does not always have the same spin polarization at the edge as the corresponding negative momentum eigenstate. In addition, in the presence of a magnetization pointing perpendicular to the edge, an edge charge current is produced, which can be chiral or nonchiral depending on whether the magnetization lies in-plane or out-of-plane. The spin polarization near the edge develops a transverse component orthogonal to the magnetization, which is antisymmetric between the two edges and tends to cause a noncollinear magnetic order between the two edges. If the magnetization only occupies a region near one edge, or in an irregular shaped quantum dot, this transverse component has a nonzero average, rendering a gate voltage-induced magnetoelectric torque without the need of a bias voltage. We also argue that other types of spin-orbit coupling that can be obtained from the Rashba type through a unitary transformation, such as the Dresselhaus spin-orbit coupling, will have similar effects too.

\end{abstract}

%\pacs{85.75.-d, 72.25.Pn, 75.85.+t}

% 85.75.-d Magnetoelectronics; spintronics: devices exploiting spin polarized transport or integrated magnetic fields
% 72.25.Pn Current-driven spin pumping

% 75.85.+t Magnetoelectric effects, multiferroics

\maketitle

\section{Introduction}

The application to information technology has been a major motivation for the development of spintronic devices. Along this line of progress, efforts towards miniaturization of such devices made effects of geometric confinement a natural target of investigation. In particular, the spintronics related effects at the edge of two-dimensional (2D) thin films or heterostructures are of special interest, since most envisaged devices rely on this type of design. In this context, 2D metallic thin films with Rashba spin-orbit coupling (RSOC) have been studied \cite{Bychkov84,Bychkov84_2}, in parts, for the current induced spin-orbit torque (SOT) \cite{Manchon08,Manchon09,Haney10,Gambardella11,Pesin12,Haney13} and the spin Hall effect \cite{Hirsch99,Sinova04,Kato04,Valenzuela06,Sinova15}, both of which have been demonstrated in those devices for practical applications. Notable edge effects in this kind of Rashba systems include an equilibrium charge current induced by an in-plane magnetization pointing perpendicular to the edge \cite{Usaj05}, and an out-of-plane polarized edge spin current in the presence of a magnetic field or magnetization \cite{Reynoso04,Grigoryan09}. Besides, the out-of-plane polarized spin current occurs at the interface separating regions with and without RSOC \cite{Sablikov08}. Because the RSOC can be adjusted by a gate voltage, manipulating these edge effects under experimentally accessible condition should in principle be feasible.

In a recent work \cite{deSousa21_RSOC_nanoribbon}, we have demonstrated several peculiar spintronic effects caused by RSOC near the edge of yet another major category of 2D systems, namely the graphene nanoribbons. The investigation was motivated by features of nanoribbons that distinguish them from usual 2D metallic strips, including the linear dispersion at low-energy electronic spectrum \cite{Novoselov05,Zhou06,Geim07,CastroNeto09,Geim09,DasSarma11}, the existence of zero energy edge states \cite{Fujita96,Nakada96,Brey06,Ezawa06,Peres06,Akhmerov08,Tao11,Ruffieux16}, and the unusual magnetic responses at the edge \cite{Wakabayashi99,Lee05,Pisani07}. Moreover, experimental evidence suggests the possibility of a gate-controllable RSOC in some graphene heterostructures \cite{Yang16,Wang16_3,Yang17,Safeer19,Ghiasi19,Benitez20,Mendes15,
Wang15_3,Dushenko16,Leutenantsmeyer17,Rybkin18}, pointing to a way of realizing nanoribbons with RSOC. In the present work, we further demonstrate that some of the peculiar spintronic effects found in our previous study are in fact not unique to graphene nanoribbons, but rather universal features for geometrically confined 2D RSOC systems. Using a square lattice tight-binding model to simulate usual 2D metallic strips containing RSOC {  and magnetization}, we uncover the following spintronic effects similar to that in the graphene nanoribbons.

{  The first feature is that the out-of-plane polarized helical edge spin current caused by RSOC alone, found already by Usaj and Balseiro\cite{Usaj05}, can be attributed to the fact that RSOC yields counter propagating modes for the two spin components due to a spin dependent shift of the electronic momentum.} A magnetization non-parallel to the edge then induces a charge current at the edge. In a ribbon geometry we observe situation where the charge current flows symmetric (non-chiral) or an antisymmetric (chiral) between the two ribbon edges, depending on whether the spin polarization is in-plane or out-of-plane, based on two different mechanisms. 
Additionally, a transverse spin polarization appears, which is antisymmetric and tends to cause non-collinear magnetic order between the two edges. There are, however, situations where a net spin polarization can be generated, which would exert a spin torque on the magnetization. Because RSOC can be tuned by a gate voltage in the device, this mechanism yields a magnetoelectric torque without any bias voltage, which is very different from the standard SOT. {  In addition, we provide a symmetry argument for the symmetric and antisymmetric patterns of these currents and spin polarizations between the two edges, and elaborate that they are generic phenomena that occur in any type of SOC that can obtained from the RSOC from a unitary transformation, such as the Dresselhaus SOC. }

%The first is that RSOC causes an out-of-plane polarized helical edge spin current, similar to that in the quantum spin Hall effect (QSHE), even without a magnetic field or magnetization. This helical edge spin current occurs because if the wave function of $+k_{x}$ state has spin up at one edge, then often times the corresponding $-k_{x}$ state has spin down at the same edge. In other words, there exist counterpropagating spins at the edge. In the presence of a magnetization pointing perpendicular to the edge, an edge charge current is produced. Although this phenomenon has been reported previously at a single edge, we point out that this current can be either symmetric (nonchiral) or antisymmetric (chiral) between the two edges depending on the magnetization lies in-plane or out-of-plane, each produces the edge current based on a different mechanism. Finally, a transverse spin polarization orthogonal to the magnetization is produced at the edge, which is antisymmetric and hence tends to produce a noncollinear magnetic order between the two edges. We further suggest few situations in which this transverse spin polarization averages to finite, which would exert a spin torque on the magnetization. Since the RSOC can be tuned by an out-of-plane electric field, this indicates a gate-induced magnetoelectric torque without applying any bias voltage, a mechanism that is entirely different from the prototype spin-orbit torque. 

We structure this paper by introducing in Sec.~\ref{sec:lattice_model} our lattice model and the corresponding current operators, and then turn to Sec.~\ref{sec:Rashba_ladder} to solve the 2-leg ladder version of our model to demonstrate analytically the existence of edge currents and spin currents, as well as the existence of transverse spin polarization. In Sec.~\ref{sec:Rashba_strip}, we numerically demonstrate these phenomena for our model defined on a strip of finite width. Section \ref{sec:finite_torque} discusses several designs that can render a finite spin torque for practical applications. These results are then summarized in Sec.~\ref{sec:conclusions}.

\section{2D Rashba magnets}

\subsection{Lattice model of 2D magnets with RSOC \label{sec:lattice_model}}

The system under consideration is a 2D square lattice which we describe by tight-binding model including RSOC and a term controlling the spin polarization, {  
\begin{eqnarray}
&&H=H_{t}+H_{J}+H_{RSOC},
\nonumber \\
&&H_{t}=-t\sum_{\langle ij\rangle \sigma}c_{i \sigma }^{\dag}c_{j\sigma} -\mu\sum_{i\sigma}c_{i\sigma}^{\dag}c_{i\sigma},
\nonumber \\
&&H_{J}=J_{ex}\sum_{i,\alpha,\beta}{\bf S}\cdot c_{i\alpha}^{\dag}{\boldsymbol \sigma}_{\alpha\beta}c_{i\beta},
\nonumber \\
&&H_{RSOC}=i\lambda_{R}\sum_{\langle ij\rangle,\alpha,\beta}c_{i\alpha}^{\dag}({\boldsymbol\sigma}_{\alpha\beta}\times{\hat{\bf d}}_{ij})^{z}c_{j\beta},
\label{By_SOC_lattice_Hamiltonian}
\end{eqnarray} }
where $c_{i\sigma}$ is the annihilation operator of an electron at site $i=(x,y)$ and with spin $\sigma$, %$\delta=(a,b)$ are the lattice constants, 
$t $ is the hopping matrix element between neighboring sites $\langle ij\rangle$ which we take as the energy unit, $\mu$ is the chemical potential, and ${\bf S}=(S^{x},S^{y},S^{z})=S(\sin\theta\cos\phi,\sin\theta\sin\phi,\cos\theta)$ is a magnetization that couples locally to the spin via an exchange coupling $J_{ex}$. The RSOC has the coupling strength $\lambda_{R}$ and ${\hat {\bf d}}_{ij}=\left\{\pm{\hat{\bf x}},\pm{\hat{\bf y}}\right\}$ is the directed unit vector connecting site $i$ to the nearest-neighbor site $j$. {  Throughout the paper we use the parameter values $\lambda_{R}/t=J_{ex}|{\bf S}|/t=0.2$. Assuming the absolute scale of the hopping amplitude to be $t\sim$eV, these parameters are of the same order of magnitude as several realistic materials: The exchange coupling in typical ferromagnets such as Co, Fe, or Ni is estimated to be $J_{ex}|{\bf S}|\sim 0.01$ to $0.1$eV\cite{Liechtenstein87,Pajda01}. The RSOC varies significantly in different materials, but can be as large as $\lambda_{R}\sim 0.1$eV in Au\cite{LaShell96,Hoesch04}, Bi\cite{Koroteev04}, Bi/Ag alloy\cite{Ast07,He08,Frantzeskakis08}, and the surface states of Bi$_{2}$Se$_{3}$\cite{King11,Zhu11}. We aim at simulating systems with large $J_{ex}|{\bf S}|$ and $\lambda_{R}$ to better visualizing the spin-dependent momentum shift, as pointed out in the following sections. }

%for the sake of visualizing the effect on the band structure, but the phenomena we uncovered are certainly present at smaller and more realistic parameters. 

We define the current operators from the time-evolution of the charge $n_{i}=\sum_{\sigma}c_{i\sigma}^{\dag}c_{i\sigma}$ and spin operators $m_{i}^{\nu}= \sum_{\alpha,\beta} c_{i\alpha}^{\dag}\sigma_{\alpha\beta}^{\nu}c_{i\beta}$ at site $i$, which can be rewritten as \cite{Zegarra20,Chen20_TIFMM}
\begin{eqnarray}
\dot{n}_{i}&=&\frac{i}{\hbar}\left[H,n_{i}\right]=\frac{i}{\hbar}\left[H_{t}+H_{R},n_{i}\right]
\nonumber \\
&=&-{\boldsymbol\nabla}\cdot{\bf J}_{i}^{0}=-\frac{1}{a}\sum_{\delta}J_{i,i+\delta}^{0},
\nonumber \\
\dot{m}_{i}^{\nu}&=&\frac{i}{\hbar}\left[H,m_{i}^{\nu}\right]=\frac{i}{\hbar}\left[H_{t}+H_{R},m_{i}^{\nu}\right]
+\frac{i}{\hbar}\left[H_{J},m_{i}^{\nu}\right]
\nonumber \\
&=&-{\boldsymbol\nabla}\cdot{\bf J}_{i}^{\nu}+\tau_{i}^{\nu}
\nonumber \\
&=&-\frac{1}{a}\sum_{\delta}J_{i,i+\delta}^{\nu}+\frac{2J_{ex}}{\hbar} \sum_{\alpha,\beta} \left({\bf S}\times c_{i\alpha}^{\dag} {\boldsymbol\sigma}_{\alpha\beta}c_{i\beta}\right)^{\nu},
\label{Mdot_1st}
\end{eqnarray}
where $\tau_{i}^{\nu}$ is interpreted as the spin torque that the magnetization excerts on the spin. $J_{i,i+\delta}^{\nu}$ denotes the charge and spin currents running from site $i$ to $i+\delta$, polarized along $\sigma^{\nu}=\left\{\sigma^{0},\sigma^{x},\sigma^{y},\sigma^{z}\right\}$, where $\delta=\left\{a,b\right\}$ represent the hopping directions along the two planar directions $ \{ \hat{\bf x}, \hat{\bf y} \}$, respectively . The resulting current operators along $a$ and $b$ directions are
\begin{eqnarray}
J_{i,i\pm a}^{\nu}&=&\frac{ia\,t}{\hbar} \sum_{\alpha,\beta} \left\{c_{i\pm a\alpha}^{\dag}\sigma_{\alpha\beta}^{\nu}c_{i\beta}
- c_{i\alpha}^{\dag}\sigma_{\alpha\beta}^{\nu}c_{i\pm a\beta}\right\}
\nonumber \\
&&\mp \frac{ia\lambda_{R}}{\hbar}\left\{
\sum_{\alpha} \left( c_{i\alpha}^{\dag}\sigma_{\alpha\uparrow}^{\nu}c_{i\pm a\downarrow} - c_{i\alpha}^{\dag}\sigma_{\alpha\downarrow}^{\nu}c_{i\pm a\uparrow} \right)\right.
\nonumber \\
&&\left. + \sum_{\beta} \left( c_{i\pm a\uparrow}^{\dag}\sigma_{\downarrow\beta}^{\nu}c_{i\beta}
- c_{i\pm a\downarrow}^{\dag}\sigma_{\uparrow\beta}^{\nu}c_{i\beta} \right) \right\},
\end{eqnarray}
\begin{eqnarray}
J_{i,i\pm b}^{\nu}&=&\frac{ib\,t}{\hbar} \sum_{\alpha,\beta} \left\{c_{i\pm b\alpha}^{\dag}\sigma_{\alpha\beta}^{\nu}c_{i\beta}
- c_{i\alpha}^{\dag}\sigma_{\alpha\beta}^{\nu}c_{i\pm b\beta}\right\}
\nonumber \\
&& \mp \frac{b\lambda_{R}}{\hbar}\left\{\sum_{\alpha} \left( c_{i\alpha}^{\dag}\sigma_{\alpha\uparrow}^{\nu}c_{i\pm b\downarrow} +c_{i\alpha}^{\dag}\sigma_{\alpha\downarrow}^{\nu}c_{i\pm b\uparrow} \right)    \right.
\nonumber \\
&&\left.+  \sum_{\beta} \left( c_{i\pm b\uparrow}^{\dag}\sigma_{\downarrow\beta}^{\nu}c_{i\beta}
 + c_{i\pm b\downarrow}^{\dag}\sigma_{\uparrow\beta}^{\nu}c_{i\beta} \right) \right\}   .
\end{eqnarray}
In general, $J_{i,i+\delta}^{\nu}$ and $J_{i+\delta,i}^{\nu}$ are not equal in the presence of boundaries. Thus, we use the following combinations to define the $\sigma^{\nu}$ component of the charge and spin current operators at site $i$ flowing along $\delta$ direction
\begin{eqnarray}
J_{\delta}^{\nu}(i)=\frac{1}{2}\left(J_{i,i+\delta}^{\nu}-J_{i+\delta,i}^{\nu}\right),
\label{lattice_spin_current_operator}
\end{eqnarray}
where the subscript denotes the direction of flow  and the superscript the spin polarization. In Appendix \ref{app:continuous_limit}, we demonstrate explicitly that in the continuous and small momentum limit, the Hamiltonian in Eq.~(\ref{By_SOC_lattice_Hamiltonian}) and the current operators in Eq.~(\ref{lattice_spin_current_operator}) recover the usual expressions of RSOC systems with parabolic bands. Finally, having the current operators defined, the equilibrium charge and spin currents are calculated by the expectation values $\langle J_{\delta}^{\nu}(i)\rangle=\sum_{n}\langle n|J_{\delta}^{\nu}(i)|n\rangle f(E_{n})$, where $|n\rangle$ is the $n$-th eigenstate with energy  $E_{n}$ for the lattice Hamiltonian, and $f(E_{n})=(e^{E_{n}/k_{B}T}+1)^{-1}$ is the Fermi distribution function. The bracket of the expectation value $\langle{\hat{\cal O}}\rangle\equiv {\hat{\cal O}}$ is often omitted in the discussion for simplicity. We have also verified that the expectation value of the continuity equation defined according to Eq.~(\ref{Mdot_1st}) vanishes on every site $\langle\dot{m}_{i}^{\nu}\rangle=\langle\dot{n}_{i}\rangle=0$ regardless the parameters and boundary conditions, such that the system remains at equilibrium.

\subsection{2-leg Rashba ladder \label{sec:Rashba_ladder}}

\begin{figure}[ht]
\begin{center}
\includegraphics[clip=true,width=0.99\columnwidth]{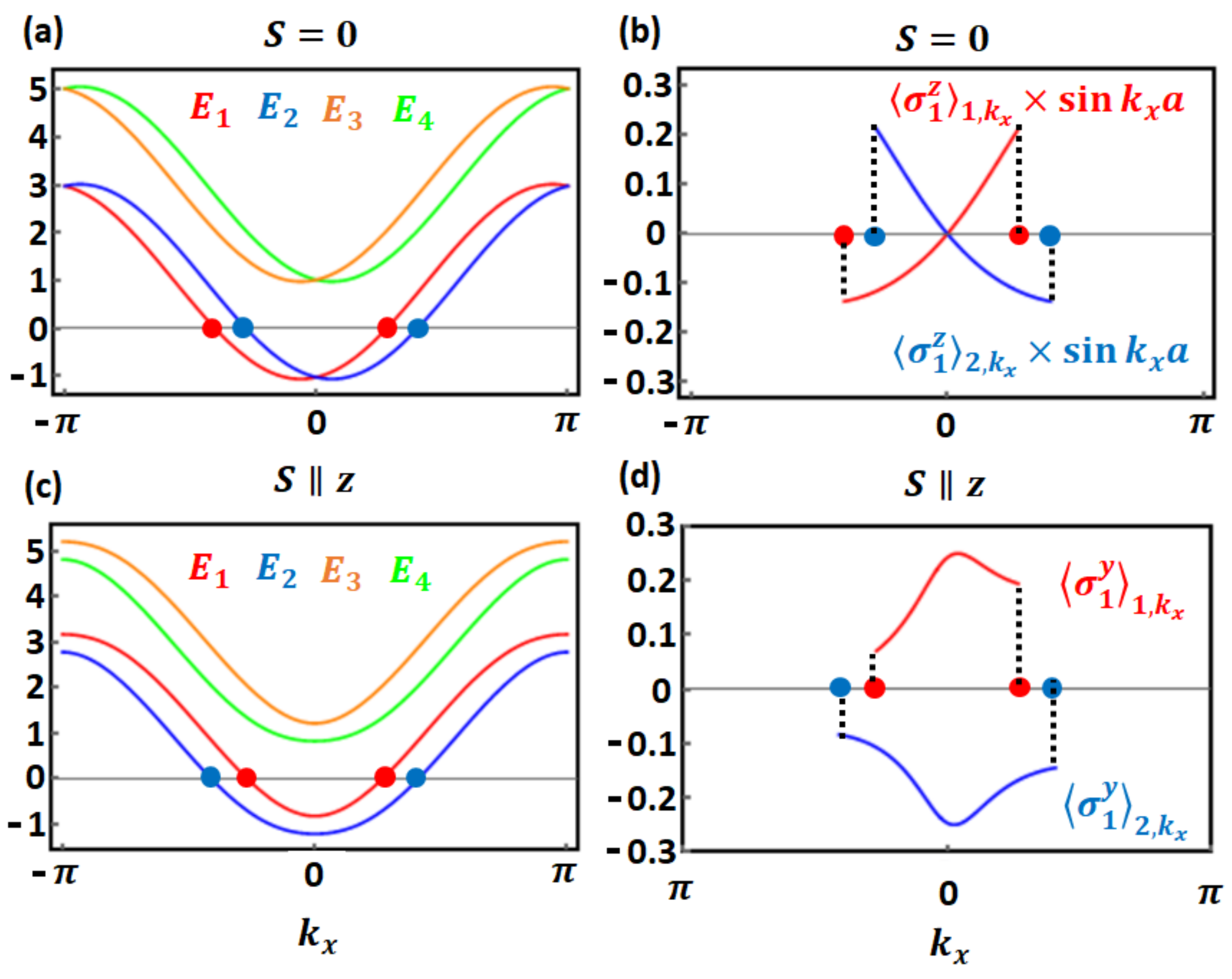}
\caption{(a) The band structure of the 2-leg Rashba ladder in the absence of magnetization ${\bf S=0}$, where the colored points indicate the Fermi momenta assuming only the two lowest bands are filled. (b) The out-of-plane spin polarization on leg 1 contributed from filled eigenstates of the first $\langle\sigma^{z}_{1}\rangle_{1,k_{x}}$ and the second $\langle\sigma^{z}_{1}\rangle_{2,k_{x}}$ band multiplied by $\sin k_{x}a$. The integrations of these two curves are finite, yielding the helical edge spin current. (c) The band structure in the presence of an out-of-plane magnetization ${\bf S}\parallel{\hat{\bf z}}$, and (d) the transverse spin polarization on leg 1 contributed from the two bands $\langle\sigma^{y}_{1}\rangle_{1,k_{x}}$ and $\langle\sigma^{y}_{1}\rangle_{2,k_{x}}$, which integrate to finite. } 
\label{fig:2leg_ladder_Ek_sigma}
\end{center}
\end{figure}

Before addressing a general 2D Rashba strip, we first examine the most narrow strip, a 2-leg ladder, to demonstrate analytically the realization of helical spin currents and the transverse spin susceptibility in equilibrium. For this purpose we adapt the model in Eq.~(\ref{By_SOC_lattice_Hamiltonian}) with a 
2-leg ladder lattice structure which extended along ${\hat{\bf x}}$ direction. In particular, for this ladder we denote the electron operator as $c_{Ii\sigma}$ 
with the leg and rung index, $I=\left\{1,2\right\} $ and $i=\left\{1....N\right\}$, respectively, and the spin $ \sigma $. We then introduce the four-component spinor $\psi_{k_x} =\left(c_{1k_{x}\uparrow},c_{2k_{x}\uparrow},c_{1k_{x}\downarrow},c_{2k_{x}\downarrow}\right)$ with $c_{Ii\sigma}=\sum_{k_{x}}e^{ik_{x}r_{i}}c_{Ik_{x}\sigma}/\sqrt{N}$, defining the momentum $ k_x $ along the ladder. In this way we can rewrite the Hamiltonian in terms of a $4 \times 4$-matrix $H = \sum_{k_x} \psi_{k_x}^{\dag} H(k_x) \psi_{k_x}$, where we use
\begin{widetext}
\begin{eqnarray}
&&H(k_{x})=\left(\begin{array}{cccc}
\xi_{k}+J_{ex}S^{z} & -t & -2i\lambda_{R}X_{k}+J_{ex}S_{\perp}e^{-i\varphi} & i\lambda_{R} \\
-t & \xi_{k}+J_{ex}S^{z} & -i\lambda_{R} & -2i\lambda_{R}X_{k}+J_{ex}S_{\perp}e^{-i\varphi} \\
2i\lambda_{R}X_{k}+J_{ex}S_{\perp}e^{i\varphi} & i\lambda_{R} & \xi_{k}-J_{ex}S^{z} & -t \\
-i\lambda_{R} & 2i\lambda_{R}X_{k}+J_{ex}S_{\perp}e^{i\varphi} & -t & \xi_{k}-J_{ex}S^{z}
\end{array}\right),
\end{eqnarray}
with 
 \begin{equation}
 \xi_{k}=-2t\cos k_{x}a-\mu,\;\;\;X_{k}=\sin k_{x}a,\;\;\;S_{\perp}=\sqrt{(S^{x})^{2}+(S^{y})^{2}}.
\end{equation}
\end{widetext}
Let us first discuss the situation without magnetization $({\bf S}={\bf 0})$, to analyze the effect of RSOC alone. Diagonalizing the Hamiltonian yields the eigenenergies 
\begin{eqnarray}
&&E_{n_y}^{{\bf S=0}}(k_{x})=\xi_{k}\pm Z_{\pm k},
\nonumber \\
&&Z_{k\pm}=\left(t^{2}\pm 4X_{k}t\lambda_{R}+\lambda_{R}^{2}+4X_{k}^{2}\lambda_{R}^{2}\right)^{1/2},
\label{2leg_S0_Ek}
\end{eqnarray}
which are displayed in Fig.~\ref{fig:2leg_ladder_Ek_sigma} (a). {  Here the eigenstates $|u_{n_{y},k_{x}}\rangle$ and eigenenergies $E_{n_y}(k_{x})$ are labeled by the momentum $k_{x}$ and the band index $n_y =\left\{1, \dots, 4\right\}$. The band structure in Fig.~\ref{fig:2leg_ladder_Ek_sigma} (a)} can be easily understood: The usual 1D metallic $\cos k_{x}a$ band is split into two due to RSOC, one shifted by a positive and the other by a negative momentum. On top of this, the rung hopping and RSOC yield a splitting into high- and a low-energy bands corresponding to the anti-bonding and bonding configurations, respectively.

For the eigenstates $|u_{n_{y},k_{x}}\rangle$, we determine the out-of-plane spin polarization as $\langle u_{n_{y},k_{x}}|\sigma^{z}_{I}|u_{n_{y},k_{x}}\rangle\equiv\langle\sigma^{z}_{I}\rangle_{n_{y},k_{x}}$ on leg $I$: 
\begin{eqnarray}
&&\langle\sigma^{z}_{1}\rangle_{1,k_{x}}=-\langle\sigma^{z}_{2}\rangle_{1,k_{x}}
=-\langle\sigma^{z}_{1}\rangle_{2,k_{x}}=\langle\sigma^{z}_{2}\rangle_{2,k_{x}}=\frac{\lambda_{R}}{2Z_{k-}},
\nonumber \\
&&-\langle\sigma^{z}_{1}\rangle_{3,k_{x}}=\langle\sigma^{z}_{2}\rangle_{3,k_{x}}
=\langle\sigma^{z}_{1}\rangle_{4,k_{x}}=-\langle\sigma^{z}_{2}\rangle_{4,k_{x}}=\frac{\lambda_{R}}{2Z_{k+}}.
\nonumber \\
\label{2leg_spin_expect}
\end{eqnarray}
The form of $Z_{k\pm}$ in Eq.~(\ref{2leg_S0_Ek}) indicates that the spin polarization is neither symmetric nor antisymmetric under reversal of $k_x$, i.e.: 
\begin{eqnarray}
\langle\sigma_{I}^{z}\rangle_{n_{y},k_{x}}\neq\pm\langle\sigma_{I}^{z}\rangle_{n_{y},-k_{x}}.
\label{S0_sigmaz_symmetry}
\end{eqnarray}
According to the last equation in Eq.~(\ref{current_momentum_space}), the out-of-plane polarized spin current $ J_{x,I}^z $, on the two legs $I=1,2$ at zero temperature is given by
\begin{eqnarray}
&&J_{x,1}^{z}=-J_{x,2}^{z}
\nonumber \\
&&=\frac{2ta}{\hbar}\sum_{n_{y}}\int_{k_{F,n_{y}}^{-}}^{k_{F,n_{y}}^{+}}\frac{dk_{x}}{2\pi}\langle\sigma^{z}_{1}\rangle_{n_{y},k_{x}}\sin k_{x}a,
\label{2leg_Jz1_Jz2}
\end{eqnarray}
where $ k_{F,n_{y}}^{\pm} $ are the two Fermi points of each band $ n_y $ with the superscript $ +,- $ indicating the right- and left-hand-side Fermi momentum ($ k_{F,n_{y}}^{+} > k_{F,n_{y}}^{-} $), as indicated by the colored points in Fig.~\ref{fig:2leg_ladder_Ek_sigma}. This corresponds to integration of the function $\left(\lambda_{R}/2Z_{k\pm}\right)\times\sin k_{x}a$ over all the occupied states according to Eq.~(\ref{2leg_spin_expect}), yielding a nonzero value. Note that for completely filled bands, $ k_{F,n_{y}}^{\pm} = \pm \pi $ yielding a zero contribution. 
Fig.~\ref{fig:2leg_ladder_Ek_sigma} (b) shows the integrand $\langle\sigma^{z}_{1}\rangle_{n_{y},k_{x}}\sin k_{x}a$ in Eq.~(\ref{2leg_Jz1_Jz2}) for the two lowest bands $E_{1}$ and $E_{2}$ used in the panel (a), which are partially filled. 
The two legs possess opposite currents $J_{x,1}^{z}=-J_{x,2}^{z}$ and represent, thus, a helical edge spin current.

%The current entirely originates from RSOC, as can be seen from the $\lambda_{R}$-dependence in Eq.~(\ref{2leg_spin_expect}). 

Turning on the magnetization ${\bf S}\parallel{\hat{\bf z}}$, the eigenenergies  become
\begin{eqnarray}
&&E_{n_y}^{\bf S\parallel{\hat z}}(k_{x})=\xi_{k}\pm\sqrt{t^{2}+\lambda_{R}^{2}+Y_{k}
\pm\sqrt{t^{2}Y_{k}}},
\nonumber \\
&&Y_{k}=J_{ex}^{2}S^{2}+4\lambda_{R}^{2}X_{k}^{2},
\end{eqnarray}
as displayed in Fig.~\ref{fig:2leg_ladder_Ek_sigma} (c). A key feature of this case is the non-vanishing $\langle\sigma_{I}^{y}\rangle_{n_{y},k_{x}}$, orthogonal to ${\bf S}\parallel{\hat{\bf z}}$, whose analytical expression is omitted here, as it is rather lengthy. Nevertheless, $\langle\sigma_{I}^{y}\rangle_{n_{y},k_{x}}$ is neither symmetric nor antisymmetric under inversion of momentum 
\begin{eqnarray}
\langle\sigma_{I}^{y}\rangle_{n_{y},k_{x}}\neq\pm\langle\sigma_{I}^{y}\rangle_{n_{y},-k_{x}},
\label{Salongz_sigmay_symmetry}
\end{eqnarray}
as shown in Fig.~\ref{fig:2leg_ladder_Ek_sigma} (d) for the two partially filled bands. The presence of RSOC is essential as the {  spin polarization} vanishes otherwise, and it is antisymmetric between the two legs, 
\begin{eqnarray}
\langle\sigma_{1}^{y}\rangle_{n_{y},k_{x}}=-\langle\sigma_{2}^{y}\rangle_{n_{y},-k_{x}}.
\end{eqnarray}
As a result, the partially filled bands contribute to a net transverse spin poalrization orthogonal to the $ {\bf S} $ and opposite in sign between the two legs, as given by
\begin{eqnarray}
\langle\sigma^{y}_{1}\rangle=-\langle\sigma^{y}_{2}\rangle=\sum_{n_{y}}
\int_{k_{F,n_{y}}^{-}}^{k_{F,n_{y}}^{+}}\frac{dk_{x}}{2\pi}\langle\sigma_{1}^{y}\rangle_{n_{y},k_{x}}
\end{eqnarray}
Moreover, according to Eq.~(\ref{current_momentum_space}), the charge current flowing in the ladder direction $J_{x}^{0}$ contains two terms. The first term is the usual hopping term calculated by the integration of $\propto\langle\sigma_{I}^{0}\rangle_{n_{y},k_{x}}\sin k_{x}a$, which vanishes because $\langle\sigma_{I}^{0}\rangle_{n_{y},k_{x}}=1/2$ and $\sum_{n_y} \int_{k_{F,n_{y}}^{+}}^{k_{F,n_{y}}^{-}}\sin k_{x}a\,dk_{x}=0$, as can be easily seen considering Fig.~\ref{fig:2leg_ladder_Ek_sigma} (c). However, the second term contributes to a finite current
\begin{eqnarray}
&&J_{x,1}^{0}=-J_{x,2}^{0}
\nonumber \\
&&=\frac{2\lambda_{R}a}{\hbar}\sum_{n_{y}}
\int_{k_{F,n_{y}}^{-}}^{k_{F,n_{y}}^{+}}\frac{dk_{x}}{2\pi}
\langle\sigma_{1}^{y}\rangle_{n_{y},k_{x}}\cos k_{x}a,
\end{eqnarray}
whose flow direction is opposite between the two legs, indicating a chiral edge charge current.

In summary, this two-leg ladder model gives an analytical account for the important features we will discuss now for strips of larger width. We find a helical spin edge current with out-of-plane spin orientation caused by RSOC. Moreover, the spin polarization ${\bf S\parallel{\hat{z}}}$ induces a transverse spin polarization $\langle\sigma^{y}\rangle$ of opposite sign on the two legs, accompanied by a chiral (charge) current on the legs.

\subsection{Edge currents and magnetoelectric torque in a 2D RSOC strip \label{sec:Rashba_strip}}

\begin{figure}[ht]
\begin{center}
\includegraphics[clip=true,width=0.99\columnwidth]{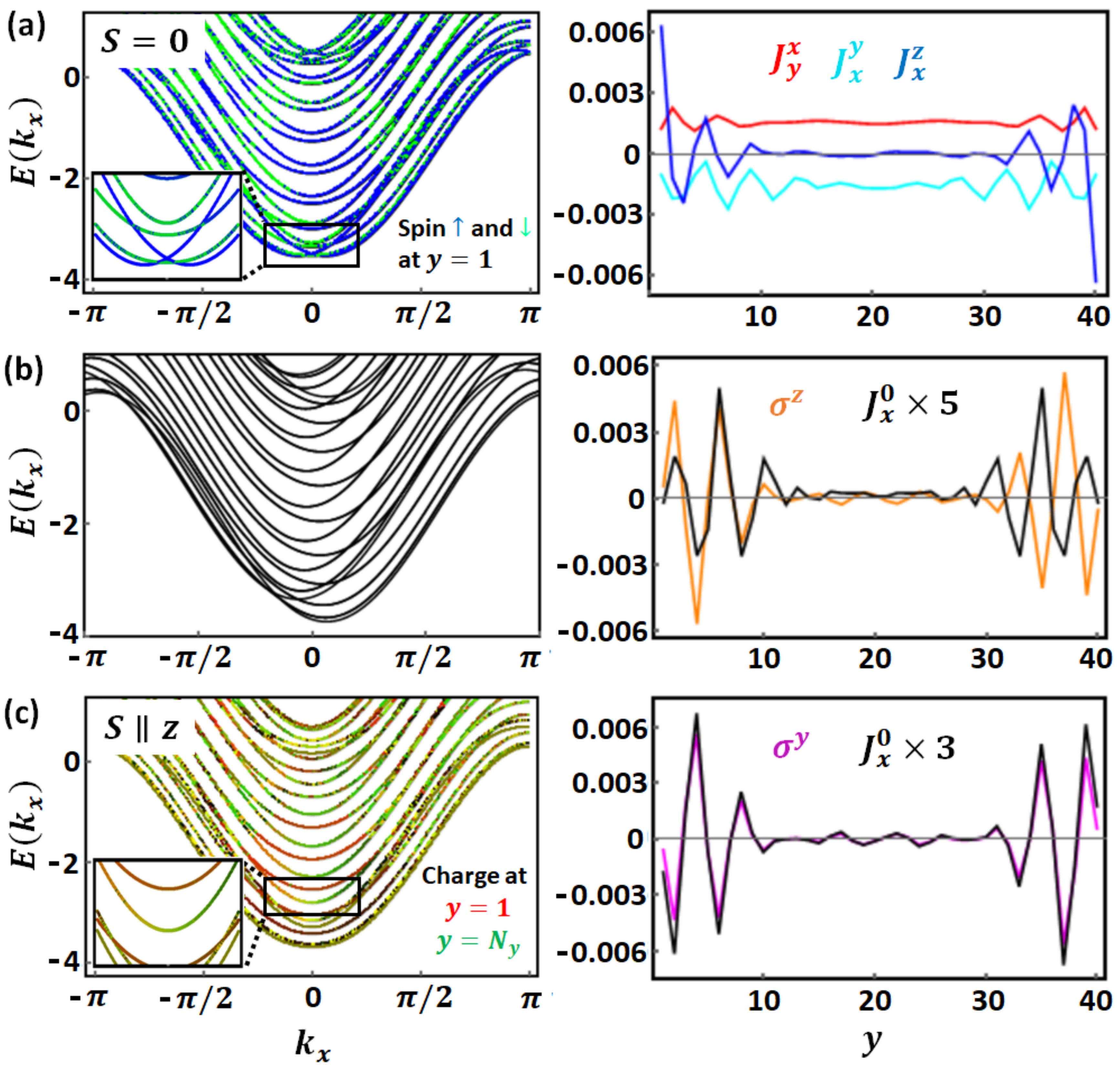}
\caption{(a) The numerical result of 2D Rashba strip in the absence of magnetization ${\bf S=0}$. The band structure (left) is colored according to the spin at the edge $y=1$ of the eigenstates, {  and the inset highlights the behavior at the bottom of the bands. The coloring of the band structure indicates that spins at momentum $+k_{x}$ and $-k_{x}$ are often opposite for some bands, suggesting an equilibrium edge spin current.} The spin current $J_{\beta}^{\alpha}$ polarized in $\alpha$-direction and flowing along $\beta$-direction is shown in the right panel, where we see the bulk spin current $J_{x}^{y}$ and $J_{y}^{x}$ due to RSOC, and in addition an out-of-plane polarized helical edge spin current $J_{x}^{z}$. (b) The ${\bf S}\parallel{\hat{\bf y}}$ situation, whose band structure is asymmetric between $+k_{x}$ and $-k_{x}$, resulting in a laminar edge current $J_{x}^{0}$ that is symmetric between the two edges. In addition, a transverse spin polarization $\langle\sigma^{z}\rangle$ is induced. (c) For the ${\bf S}\parallel{\hat{\bf z}}$ case, the band structure is colored according to the wave function at the left $y=1$ and right $y=N_{y}$ edges, {  and the inset shows a section of the band structure in more detail. The coloring of the band structure shows that $+k_{x}$ is more localized at the right edge and $-k_{x}$ more localized at the left edge for some bands, indicating a chiral edge current $J_{x}^{0}$.} In addition, a transverse spin polarization $\langle\sigma^{y}\rangle$ is induced, which almost has the same pattern as the $J_{x}^{0}$. } 
\label{fig:2DEGRashbamag_S0xyz}
\end{center}
\end{figure}

We proceed now with the Hamiltonian in Eq.~(\ref{By_SOC_lattice_Hamiltonian}) defined on a 2D strip with periodic boundary condition (PBC) along ${\hat{\bf x}}$ and open boundary condition (OBC) along ${\hat{\bf y}}$. The numerical calculation is performed with chemical potential fixed at $\mu=-0.5t$ and for different magnetization directions ${\bf S}\parallel\left\{{\hat{\bf x}},{\hat{\bf y}},{\hat{\bf z}}\right\}$. Note that at any magnetization direction, there is always a longitudinal spin polarization along the magnetization, but we will focus on the transverse spin polarization perpendicular to the magnetization since this is the component that yields the spin torque. The results are presented in Fig.~\ref{fig:2DEGRashbamag_S0xyz} for the following cases: 

(i) ${\bf S=0}$: In the absence of magnetization, the RSOC alone causes bulk equilibrium spin currents $J_{x}^{y}$ and $J_{y}^{x}$ \cite{Rashba03,Rashba05,Sonin07,Tokatly08}, as displayed in Fig.~\ref{fig:2DEGRashbamag_S0xyz} (a). In addition, we observe the out-of-plane polarized edge spin current $J_{x}^{z}$, whose magnitude decays and oscillates with distance from the edge, consistent with the previous analytical result that suggests an oscillation length set by the Fermi momentum and the decay length by the RSOC\cite{Usaj05}. While this helical edge spin current resembles the one found in quantum spin Hall systems \cite{Kane05_2,Bernevig06,Bernevig06_2,Konig07} whose flow direction depends on the chemical potential\cite{Zegarra20,Chen20_absence_edge_current}, it is not of topological origin since the bulk spectrum is not gapped in our case. The pattern of this spin current is antisymmetric between the two edges, indicating a helical edge spin current. Color-coding the band structure according to the spin of the corresponding eigenstate wave functions at $y=1$, spin up (blue) and spin down (green), we see that states with $ +k_x $ and $ -k_x $ have often opposite spin, indicating the existence of $J_{x}^{z}$. These features of $\left\{J_{x}^{y},J_{y}^{x},J_{x}^{z}\right\}$ remain valid even in the presence of magnetization $ {\bf S} $.

%{\cblue (1) Say that for the RSOC alone case, in graphene nanoribbons if $+k_{x}$ has spin up at one edge then $-k_{x}$ always has spin down at the same edge, but in 2D metallic strip this only happens sometimes for some bands.  }

%{\cblue (1) It seems like we should investigate the range of $J_{x}^{z}$ as a function of RSOC. }

(ii) ${\bf S}\parallel{\hat{\bf x}}$: For a magnetization pointing along the strip, the pattern of the spin current is practically unchanged to the case of ${\bf S=0}$. Naturally, a longitudinal spin polarization $\langle\sigma^{x}\rangle $ is induced. 
Since there are no new qualitative feature otherwise, we omit this case in Fig.~\ref{fig:2DEGRashbamag_S0xyz}.

(iii) ${\bf S}\parallel{\hat{\bf y}}$: In this case a charge current $J_{x}^{0}$ occurs close to the edge and is diminished in the center of the strip. This edge current oscillates with distance from the edge, is symmetric between the two edges and, hence, non-chiral. Its occurrence is due to a shift in the band structure, which becomes asymmetric between $+k_{x}$ and $-k_{x}$ because of the in-plane magnetization\cite{Gambardella11}.  
In addition, the spin polarization near the edge develops an out-of-plane component $\langle\sigma^{z}\rangle$, rendering a local spin torque according to the Landau-Lifshitz dynamics 
\begin{eqnarray}
\frac{d{\bf S}_{i}}{dt}=\frac{J_{ex}}{\hbar}\langle{\boldsymbol\sigma}_{i}\rangle\times{\bf S}_{i}.
\label{Landau_Lifshitz_dynamics}
\end{eqnarray}
However, because $\langle\sigma^{z}\rangle$ is antisymmetric between the two edges, the net torque is zero and, thus, does not flip the magnetization on a mesoscopic scale. Instead, the local torque has the tendency to cause non-collinear polarization between the two edges, i.e., effectively a Dzyaloshinskii-Moriya interaction (DMI) on the mesoscopic scale.

(iii) ${\bf S}\parallel{\hat{\bf z}}$: In this case, a charge edge current $J_{x}^{0}$ also occurs, but having opposite direction on the two edges indicates that the current is a chiral edge current. Combining with the ${\bf S}\parallel{\hat{\bf y}}$ case above, this suggests that the chirality of $J_{x}^{0}$ can be controlled by the direction of magnetization. The coloring of the band structure in Fig.~\ref{fig:2DEGRashbamag_S0xyz}(c) allows us to judge whether the eigenstate wave functions of given $k_x $ are localized more to the right or to left edge, {  which suggests that some bands have $+k_{x}$ more localized at the right edge and $-k_{x}$ more localized at the left edge, indicating the existence of a chiral edge current.}
%In this way the chiral charge current is reflected in the band structure. 
Moreover, a transverse spin polarization $\langle\sigma^{y}\rangle$ antisymmetric between the two edges is induced, which again yields the local torque and mesoscopic DMI according to Eq.~(\ref{Landau_Lifshitz_dynamics}) and practically retraces the spatial dependence of the charge current. 

%as expected for the inverse Edelstein effect. 

In Appendix \ref{app:symmetry_current_spin}, we demonstrate that the patterns of charge/spin currents and spin polarizations between the two edges are in agreement with the expectations from the symmetry properties of the system, in particular the mirror reflections. Comparing the three magnetization directions $ {\bf S} $, we find that the local torque can be written in the form
\begin{eqnarray}
\left.\frac{d{\bf S}_{i}}{dt}\right|_{\bf S\parallel\left\{{\hat{\bf x}},{\hat{\bf y}},{\hat{\bf z}}\right\}}\propto {\bf S}\times\left({\bf S}\times{\hat{\bf x}}\right),
\label{spin_torque_damping_like}
\end{eqnarray}
which corresponds to a damping-like torque defined with respect to the strip direction (${\hat{\bf x}}$), and is in complete agreement with the symmetry argument in Appendix \ref{app:symmetry_current_spin}. We close this section by mentioning that although we focus on Rashba systems here, in Appendix \ref{app:Dresselhaus_SOC} we demonstrate that these edge currents and transverse spin polarizations are in fact universal features for any type of antisymmetric spin-orbit coupling that can be obtained from the RSOC by a unitary transformation, such as the Dresselhaus spin-orbit coupling. Because the unitary transformation rotates the spin quantization axis, these effects occur when the magnetization is oriented accordingly in the appropriate direction.

%{\cblue (2) Well, the result seems to suggest that even without magnetization but just the Rashba SOC alone is already enough to have a dissipationless spin current induced by transverse electric polarization. }

%{\cblue (3) I should draw a spin potential (spin polarization) landscape just like I did for dissipationless charge current. Then I argue that the spin voltage from the leads can tunnel into lower spin potentials in the Rashba magnet, then cite my quantum spin tunneling paper to justify that such tunneling of spin without transferring charge is possible. }

%{\cblue (4) In principle the effect of charge polarization is the same as SHE, because one can imagine if the Rashba magnet has SHE, then a current passing in the transverse direction can also produce a spin current, and the resulting spin accumulation and so on. However, we can sell that the advantage of charge polarization is that to {\it maintain} the dissipationless spin current, it takes no energy because there is no Joule heating. However, the disadvantage compared to SHE is that the dissipationless spin current only occurs at the edge. }

\begin{figure}[ht]
\begin{center}
\includegraphics[clip=true,width=0.99\columnwidth]{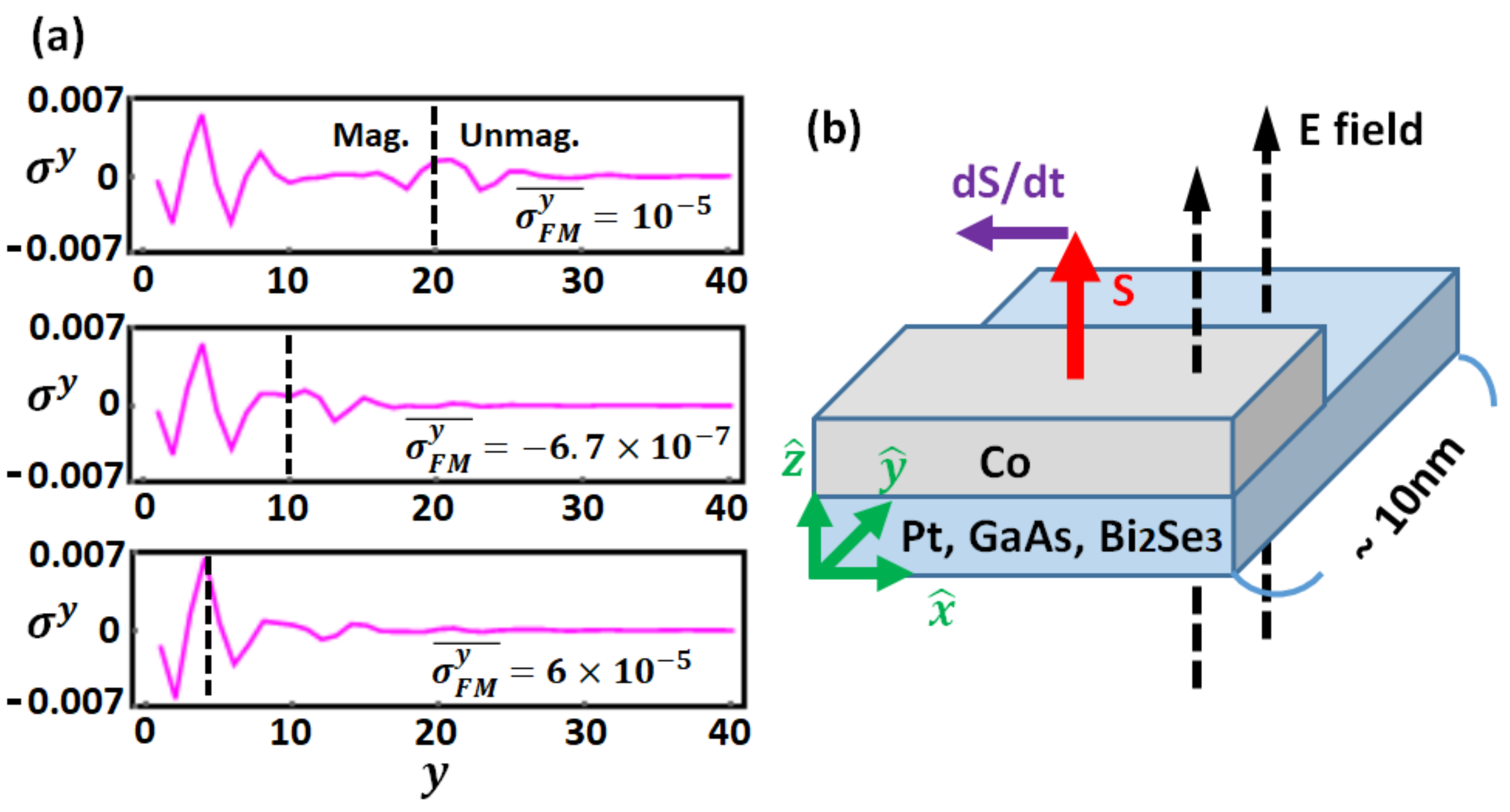}
\caption{(a) Spin polarization $\sigma^{y}$ as a function of transverse coordinate $y$ in partially magnetized RSOC strip, with the three panels represent different ranges of magnetization closer the $y=1$ edge. The averaged spin polarization in the magnetized region is indicated by $\overline{\sigma_{FM}^{y}}$. (b) A possible realization of this set up, which consists of Co deposited closer to one edge of a Pt, GaAs, {  or Bi$_{2}$Se$_{3}$ strip.} An electric field produced by a gate voltage tunes the RSOC, which is predicated to cause a magnetoelectric torque, and no bias voltage is needed. For the quantum confinement effect to take place, the width of the strip in the ${\hat{\bf y}}$ direction should be of the order of $\sim 10$nm, but the length in the ${\hat{\bf x}}$ direction is not restricted. } 
\label{fig:magrange_flake}
\end{center}
\end{figure}

\subsection{Engineering a finite magnetoelectric torque \label{sec:finite_torque}}

%{\cblue (2) How about plotting $\overline{\sigma^{y}_{FM}}$ as a function of $N_{y,RS}$ and $N_{y,FM}$, as a 3D plot?  }

To raise the magnetoelectric torque to practical usage, one must design a situation where the torque does not average to zero on a mesoscopic scale, as it happens in our simple strip above.  We suggest two situations that can resolve this problem. The first is by introducing a spatial dependent magnetization, e.g.  the exchange coupling term $H_{J}$ in Eq.~(\ref{By_SOC_lattice_Hamiltonian}) exists only for sites $i=(x,y)$ that are within a range to one of the two edges, say $y\leq N_{y,mag}$. In this situation, we can use the spin polarization averaged over this region $\overline{\sigma^{\alpha}_{FM}}=\sum_{y=1}^{N_{y,mag}}\sigma^{\alpha}_{y}/N_{y,mag}$ to quantify the torque. As shown in Fig.~\ref{fig:magrange_flake} (a) for ${\bf S}\parallel{\hat{\bf z}}$, we indeed see that in this set up, there is a finite spin polarization as large as $\overline{\sigma^{\alpha}_{FM}}\sim 10^{-5}$ in our parameters $J_{ex}|{\bf S}|=\lambda_{R}=0.2$, which would give a torque of magnitude $\sim$ GHz, comparable to that of SOT in typical devices.  
Drawing analogy with the usual SOT set up, this partially magnetized situation suggests the device shown schematically in Fig.~\ref{fig:magrange_flake} (b), where a narrow Co film is deposited closer to one edge of a thin Pt or GaAs strip. The width of the device must be narrow enough ($\sim 10$nm) for the quantum confinement effect to take place. Note that a gate voltage can efficiently tune the RSOC, as has been demonstrated in semiconductor heterostructures \cite{Schultz96,Nitta97,Nitta98,Heida98,Miller03,Hinz06,Studer09,
Koo09,Liang12,Takase17,Chen18_tunable_SOC,Froning21}, {  surface states of topological insulators Bi$_{2}$Se$_{3}$\cite{King11},} and also hinted by the experiments in heavy metal thin films \cite{Dushenko18}. Provided a magnetization can be incorporated into these devices, a magnetoelectric torque should be induced by the gate voltage without applying any bias voltage. In particular, recent experiments on Fe/GaAs\cite{Chen18_FeGaAs} and oxide-capped Pt/Co\cite{Emori14,Mishra19,Liu14_Efield_SOT} suggest the controllability of RSOC by a gate voltage in these heterostructures.

\begin{figure}[ht]
\begin{center}
\includegraphics[clip=true,width=0.99\columnwidth]{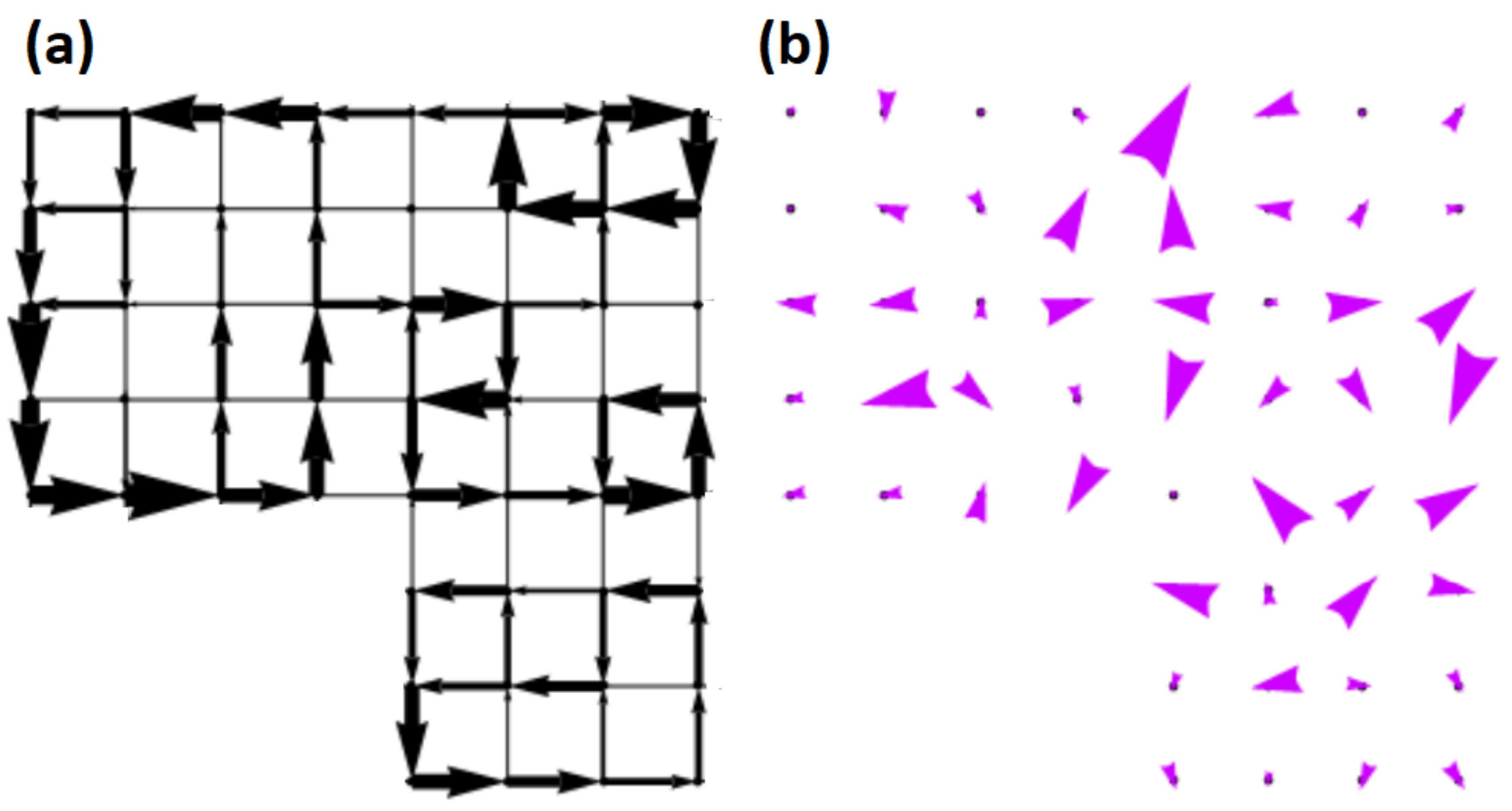}
\caption{(a) The network of equilibrium charge current $J^{0}$ in a small L-shaped RSOC nanoflake with out-of-plane magnetization ${\bf S}\parallel{\hat{\bf z}}$. The largest black arrow corresponds to current $0.039$ in units of $et/\hbar$. (b) The in-plane spin polarization in this configuration, where the largest purple arrow corresponds to spin polarization $0.041$ in units of Bohr magneton. The average in-plane spin polarization per site is $\overline{\sigma^{x}}=-8.7\times 10^{-4}$ and $\overline{\sigma^{y}}=3.6\times 10^{-4}$ in units of Bohr magneton $\mu_{B}$.  } 
\label{fig:2DEG_nanoflake}
\end{center}
\end{figure}

Another geometry that can realize a finite torque is the small, irregularly shaped nanoflake for which symmetry related cancelling effects are suppressed.  As shown in Fig.~\ref{fig:2DEG_nanoflake} (a) for an L-shaped nanoflake, the equilibrium charge current caused by the out-of-plane magnetization ${\bf S}\parallel{\hat{\bf z}}$ still exists, but turns into a complex pattern of currents that flows throughout the nanoflake. This suggests that the equilibrium charge and spin currents may occur ubiquitously in polycrystalline thin films that contain small magnetic domains or grain boundaries, in which the current pattern depends on the shape of the grain. Moreover, the in-plane spin polarization shown in Fig.~\ref{fig:2DEG_nanoflake} (b) has a nonzero average of the order of $\overline{\sigma^{x}}\sim\overline{\sigma^{y}}\sim 10^{-4}$ (in units of Bohr magneton $\mu_{B}$) per site, which would yield a very large spin torque $\sim 10$GHz, although it varies significantly with the size and shape of the nanoflake. Our result suggests that geometric confinement can play a very important role in the magnetization dynamics of nanometer scale devices, and may further be used to engineer the spin torque in these devices, provided the shape of the device is fabricated in a controlled manner.

\section{Conclusions \label{sec:conclusions}}

In summary, using a tight-binding model, we analyze a number of peculiar spintronic effects near the edge of 2D Rashba systems with strip geometry. {  Firstly, we clarify the spin-dependent momentum shift mechanism for the out-of-plane polarized helical edge spin current due to RSOC.} Secondly, introducing a magnetization perpendicular to the edge induces an edge current, which can be chiral or non-chiral depending on whether  the magnetization is oriented in-plane or out-of-plane. In addition, the spin polarization develops a component that is orthogonal to the direction of the magnetization, which serves as a local spin torque that tends to create non-collinear order between the two edges. The net torque can be finite in certain systems if the magnetization is not uniform, and it is found to be damping-like defined with respect to the edge direction as described by Eq.~(\ref{spin_torque_damping_like}). Alternatively, a nanoflake of irregular shape can also have a finite torque. Because previous investigation also unveil similar phenomena for graphene nanoribbons, this indicates that these phenomena are generic effects of the RSOC independent of lattice structure, low energy band structures, and whether localized edge states exists or not. {  Moreover, the symmetric and antisymmetric patterns of these currents and spin polarizations between the two edges can be understood from a symmetry argument.} The tunability of RSOC by a gate voltage further may allow for the controllability of these effects, which could help to observe them in reality. Finally, we elaborate that similar effects also manifest in other types of spin-orbit coupling that can be obtained from RSOC by a unitary transformation, such as the Dresselhaus spin-orbit coupling, provided the magnetization points in an appropriate direction, suggesting that these effects may occur abundantly in spin-orbit coupled systems. 

\begin{acknowledgments}
We would like to thank P. Gambardella, C. O. Avci, J. Moodera, and M. S. M. de Sousa for stimuating discussions. M.S. was financially supported by a grant of the Swiss National Science Foundation (Grant No. 184739). W. C. acknowledges the support from the productivity in research fellowship of CNPq.
\end{acknowledgments}

\appendix

\section{Continuous limit \label{app:continuous_limit}}

In this section we demonstrate explicitly that our model Hamiltonian and current operator recover the usual Rashba Hamiltonian results in the continuous and small momentum limit. Given the parametrization of a Rashba Hamiltonian in the presence of a magnetization 
\begin{eqnarray}
H=\frac{\hbar^{2}k^{2}}{2m}+\alpha_{R}\left(k_{x}\sigma^{y}-k_{y}\sigma^{x}\right)+J_{ex}{\bf S}\cdot{\boldsymbol\sigma},
\label{continuous_Rashba_Hamiltonian}
\end{eqnarray}
direct comparison with Fourier transform of Eq.~(\ref{By_SOC_lattice_Hamiltonian}) and expanding up to leading order in $(k_{x},k_{y})$ yields the identification
\begin{eqnarray}
\frac{\hbar^{2}}{2m}=a^{2}t=b^{2}t,\;\;\;\alpha_{R}=2\lambda_{R}a=2\lambda_{R}b,
\label{parameter_correspondence}
\end{eqnarray}
assuming the lattice constants in the two planar directions are the same $a=b$. The convention of defining the spin operator from the continuous Rashba model is by using the combination of velocity operator $v_{b}=\partial H/\hbar\partial k_{b}$ and the Pauli matrices\cite{Rashba03,Sun05,Jin06,Shi06,Sun08}
\begin{eqnarray}
{\cal J}_{b}^{\nu}({\bf k})=\frac{1}{2}\left(v_{b}\sigma^{\nu}+\sigma^{\nu}v_{b}\right),
\label{spin_current_definition_general}
\end{eqnarray}
where $\sigma^{\nu}=\left\{\sigma^{0},\sigma^{x},\sigma^{y},\sigma^{z}\right\}$ with $\sigma^{0}$ the identity matrix,
which yields
\begin{eqnarray}
&&{\cal J}_{x}^{0}=\frac{\hbar}{m}k_{x}\sigma^{0}+\frac{\alpha_{R}}{\hbar}\sigma^{y},
\;\;\;{\cal J}_{y}^{0}=\frac{\hbar}{m}k_{y}\sigma^{0}-\frac{\alpha_{R}}{\hbar}\sigma^{x},
\nonumber \\
&&{\cal J}_{a}^{\nu}|_{a=\left\{x,y,z\right\}}=\frac{\hbar}{m}k_{a}\sigma^{\nu},\;\;\;
{\cal J}_{x}^{y}=\frac{\hbar}{m}k_{x}\sigma^{y}+\frac{\alpha_{R}}{\hbar},
\nonumber \\
&&{\cal J}_{a}^{z}=\frac{\hbar}{m}k_{a}\sigma^{z},\;\;\;
{\cal J}_{y}^{x}=\frac{\hbar}{m}k_{y}\sigma^{x}-\frac{\alpha_{R}}{\hbar}.
\label{Rashba_current_operator}
\end{eqnarray}
On the other hand, after a Fourier transform, the sum of local spin current operator in Eq.~(\ref{lattice_spin_current_operator}) gives (omitting the electron operators $c_{k\alpha}^{\dag}c_{k\beta}$ in the expression) 
\begin{eqnarray}
&&\sum_{i}J_{x}^{0}(i)=\frac{a}{\hbar}\sum_{k}2t\sin k_{x}a\,\sigma^{0}
+\frac{a}{\hbar}\sum_{k}2\lambda_{R}\cos k_{x}a\,\sigma^{y},
\nonumber \\
&&\sum_{i}J_{y}^{0}(i)=\frac{b}{\hbar}\sum_{k}2t\sin k_{y}b\,\sigma^{0}
-\frac{b}{\hbar}\sum_{k}2\lambda_{R}\cos k_{y}b\,\sigma^{x},
\nonumber \\
&&\sum_{i}J_{a}^{\nu}(i)|_{a=\left\{x,y,z\right\}}=\frac{a}{\hbar}\sum_{k}2t\sin k_{a}a\,\sigma^{\nu},
\nonumber \\
&&\sum_{i}J_{y}^{x}(i)=\frac{b}{\hbar}\sum_{k}2t\sin k_{y}b\,\sigma^{x}
-\frac{b}{\hbar}\sum_{k}2\lambda_{R}\cos k_{y}b\,\sigma^{0},
\nonumber \\
&&\sum_{i}J_{x}^{y}(i)=\frac{a}{\hbar}\sum_{k}2t\sin k_{x}a\,\sigma^{y}
+\frac{a}{\hbar}\sum_{k}2\lambda_{R}\cos k_{x}a\,\sigma^{0},
\nonumber \\
&&\sum_{i}J_{a}^{z}(i)=\frac{a}{\hbar}\sum_{k}2t\sin k_{a}a\,\sigma^{z}.
\label{current_momentum_space}
\end{eqnarray}
Using Eq.~(\ref{parameter_correspondence}) and expanding up to leading order in $(k_{x},k_{y})$, one sees that the lattice spin current operator $\sum_{i}J_{b}^{\nu}(i)=\sum_{\bf k}{\cal J}_{b}^{\nu}({\bf k})$ agrees with the usual Rashba Hamiltonian result in Eq.~(\ref{Rashba_current_operator}), which supports the validity of Eq.~(\ref{lattice_spin_current_operator}).

\section{Symmetry argument for the currents and spin polarizations \label{app:symmetry_current_spin}}

In this section, we elaborate that the reflection symmetry along the strip direction $R_{x}$ dictates that some components of spin polarizations and current must vanish, and the reflection symmetry with respect to the middle axis of the strip $R_{y}$ determines whether the spin polarization and currents are symmetric or antisymmetric between the two edges. Under $R_{x}$, the coordinate, velocity, and spin transform as
\begin{eqnarray}
&&(x,y)\rightarrow(-x,y),
\nonumber \\
&&(v_{x},v_{y})\rightarrow(-v_{x},v_{y}),
\nonumber \\
&&(\sigma^{0},\sigma^{x},\sigma^{y},\sigma^{z})\rightarrow(\sigma^{0},\sigma^{x},-\sigma^{y},-\sigma^{z}),
\label{Rx_transformation}
\end{eqnarray}
whereas under $R_{y}$, they transform as
\begin{eqnarray}
&&(x,y)\rightarrow(x,-y),
\nonumber \\
&&(v_{x},v_{y})\rightarrow(v_{x},-v_{y}),
\nonumber \\
&&(\sigma^{0},\sigma^{x},\sigma^{y},\sigma^{z})\rightarrow(\sigma^{0},-\sigma^{x},\sigma^{y},-\sigma^{z}),
\label{Ry_transformation}
\end{eqnarray}
where $v_{\alpha}=\partial H/\hbar\partial k_{\alpha}$. The currents and spin polarizations are functions of position $(x,y)$ in the range $-N_{y}/2\leq y\leq N_{y}/2$, and can be written as
\begin{eqnarray}
&&J^{\alpha}_{\beta}(x,y)=|J^{\alpha}_{\beta}(x,y)|\,\hat{\sigma}^{\alpha}\otimes \hat{v}_{\beta},
\nonumber \\
&&\sigma^{\alpha}(x,y)=|\sigma^{\alpha}(x,y)|\,\hat{\sigma}^{\alpha},
\label{current_spin_general_form}
\end{eqnarray}
where $|J^{\alpha}_{\beta}(x,y)|$ and $|\sigma^{\alpha}(x,y)|$ denote the magnitude, $\hat{\sigma}^{\alpha}$ denotes the spin polarization, and $\hat{v}_{\beta}$ denotes the direction of flow. Because the magnitudes are scalar functions, they transform under the transverse reflection $R_{y}$ as $R_{y}|J^{\alpha}_{\beta}(x,y)|=|J^{\alpha}_{\beta}(x,-y)|$ and $R_{y}|\sigma^{\alpha}(x,y)|=|\sigma^{\alpha}(x,-y)|$, meaning that if $R_{y}$ is a symmetry of the system, then we know at least the magnitudes will be the same between $+y$ and $-y$, although their spin polarization and direction of flow may be different. The same argument also applies to the longitudinal reflection $R_{x}$. However, if a quantity $Z(x,y)$ transforms under $R_{x}$ as $R_{x}Z(x,y)=-Z(-x,y)$, then we know that $Z(x,y)$ must vanish everywhere because the translational invariance along the strip implies $Z(x,y)=Z(-x,y)$. The symmetry properties expected for the currents and spin polarization at different magnetization directions are summarized below according to whether or not $R_{x}$ and $R_{y}$ are symmetries of the Hamiltonian, which are in complete agreement with our numerical results. Moreover, the cross product of ${\bf S}$ with nonzero $\sigma^{a}$ components in each case renders the damping-like torque in Eq.~(\ref{spin_torque_damping_like}).

(i) ${\bf S}={\bf 0}$: In the absence of the magnetization, both $R_{x}$ and $R_{y}$ are symmetries for the Rashba Hamiltonian in Eq.~(\ref{continuous_Rashba_Hamiltonian}). Following how $\hat{\sigma}^{\alpha}$ and $\hat{v}_{\beta}$ transform under $R_{y}$ as given by Eq.~(\ref{Ry_transformation}) and using $R_{y}|J^{\alpha}_{\beta}(y)|=|J^{\alpha}_{\beta}(-y)|$, and likewisely for $R_{x}$, the permitted spin currents transform like
\begin{eqnarray}
&&R_{x}J_{x}^{y}(x,y)=J_{x}^{y}(-x,y),\;\;\;R_{y}J_{x}^{y}(x,y)=J_{x}^{y}(x,-y),
\nonumber \\
&&R_{x}J_{y}^{x}(x,y)=J_{y}^{x}(-x,y),\;\;\;R_{y}J_{y}^{x}(x,y)=J_{y}^{x}(x,-y),
\nonumber \\
&&R_{x}J_{x}^{z}(x,y)=J_{x}^{z}(-x,y),\;\;\;R_{y}J_{x}^{z}(x,y)=-J_{x}^{z}(x,-y),
\nonumber \\
\end{eqnarray}
meaning that the bulk spin current $J_{x}^{y}$ and $J_{y}^{x}$ are symmetric but the out-of-plane polarized $J_{x}^{z}$ must be antisymmetric between the two edges, as we have found. As a counterexample, $R_{x}J_{x}^{x}(x,y)=-J_{x}^{x}(-x,y)$ and $R_{x}J_{y}^{y,z}(x,y)=-J_{y}^{y,z}(-x,y)$ implies that $\left\{J_{x}^{x},J_{y}^{y},J_{y}^{z}\right\}$ do not exist, as is known in 2D systems with RSOC. 

(ii) ${\bf S}\parallel{\hat{\bf x}}$: In this situation, $R_{y}$ is no longer a symmetry of Eq.~(\ref{continuous_Rashba_Hamiltonian}) because of the exchange term $J_{ex}S\sigma^{x}$, but $R_{x}$ is still a symmetry, and hence 
\begin{eqnarray}
&&R_{x}J_{x}^{0}(x,y)=-J_{x}^{0}(-x,y),\;\;\;R_{y}J_{x}^{0}(x,y)\neq J_{x}^{0}(x,-y),
\nonumber \\
&&R_{x}\sigma^{x}(x,y)=\sigma^{x}(-x,y),\;\;\;R_{y}\sigma^{x}(x,y)\neq-\sigma^{x}(x,-y),
\nonumber \\
&&R_{x}\sigma^{y}(x,y)=-\sigma^{y}(-x,y),\;\;\;R_{y}\sigma^{y}(x,y)\neq\sigma^{y}(x,-y),
\nonumber \\
&&R_{x}\sigma^{z}(x,y)=-\sigma^{z}(-x,y),\;\;\;R_{y}\sigma^{z}(x,y)\neq-\sigma^{z}(x,-y),
\nonumber \\
\label{Ry_spin_Salongx}
\end{eqnarray}
{  so we know that $\left\{J_{x}^{0},\sigma^{y},\sigma^{z}\right\}$ do not exist, and $\sigma^{x}$ is symmetric between the two edges. }

(iii) ${\bf S}\parallel{\hat{\bf y}}$: Because $R_{y}$ is a symmetry of Eq.~(\ref{continuous_Rashba_Hamiltonian}) but $R_{x}$ is not, the charge current and spin polarization transform like 
\begin{eqnarray}
&&R_{x}J_{x}^{0}(x,y)\neq -J_{x}^{0}(-x,y),\;\;\;R_{y}J_{x}^{0}(x,y)=J_{x}^{0}(x,-y),
\nonumber \\
&&R_{x}\sigma^{x}(x,y)\neq\sigma^{x}(-x,y),\;\;\;R_{y}\sigma^{x}(x,y)=-\sigma^{x}(x,-y),
\nonumber \\
&&R_{x}\sigma^{y}(x,y)\neq-\sigma^{y}(-x,y),\;\;\;R_{y}\sigma^{y}(x,y)=\sigma^{y}(x,-y),
\nonumber \\
&&R_{x}\sigma^{z}(x,y)\neq-\sigma^{z}(-x,y),\;\;\;R_{y}\sigma^{z}(x,y)=-\sigma^{z}(x,-y),
\nonumber \\
\end{eqnarray}
indicating that the charge current $J_{x}^{0}$ (nonchiral) and $\sigma^{y}$ are symmetric but $\sigma^{z}$ are antisymmetric between the two edges, and $\sigma^{x}$ does not exist.

(iv) ${\bf S}\parallel{\hat{\bf z}}$: Since both $R_{x}$ and $R_{y}$ are not symmetries of Eq.~(\ref{continuous_Rashba_Hamiltonian}) due to the exchange term $J_{ex}S\sigma^{z}$, we have
\begin{eqnarray}
&&R_{x}J_{x}^{0}(x,y)\neq-J_{x}^{0}(-x,y),\;\;\;R_{y}J_{x}^{0}(x,y)\neq J_{x}^{0}(x,-y),
\nonumber \\
&&R_{x}\sigma^{x}(x,y)\neq\sigma^{x}(-x,y),\;\;\;R_{y}\sigma^{x}(x,y)\neq-\sigma^{x}(x,-y),
\nonumber \\
&&R_{x}\sigma^{y}(x,y)\neq-\sigma^{y}(-x,y),\;\;\;R_{y}\sigma^{y}(x,y)\neq\sigma^{y}(x,-y),
\nonumber \\
&&R_{x}\sigma^{z}(x,y)\neq-\sigma^{z}(-x,y),\;\;\;R_{y}\sigma^{z}(x,y)\neq-\sigma^{z}(x,-y),
\nonumber \\
\end{eqnarray}
implying that the charge current cannot be symmetric between the two edges, consistent with the chiral edge current we uncovered.

\section{Dresselhaus systems with magnetization \label{app:Dresselhaus_SOC}}

Besides the RSOC, Dresselhaus spin-orbit coupling is another common mechanism for spin-momentum locking in 2D materials. In this section, we elaborate that the laminar edge currents and transverse susceptibility also exist in the Dresselhaus systems, although the magnetization must be oriented in a different direction. Consider the 2D model
\begin{eqnarray}
H_{D}=\frac{\hbar^{2}k^{2}}{2m}+\alpha_{D}\left(k_{x}\sigma^{x}-k_{y}\sigma^{y}\right)+J_{ex}{\bf S}\cdot{\boldsymbol\sigma},
\label{Dresseihaus_continuous_model}
\end{eqnarray}
where $\alpha_{D}$ represents the strength of the Dresselhaus spin-orbit coupling. Our statement is made based on the observation that the Rashba magnet in Eq.~(\ref{continuous_Rashba_Hamiltonian}) and the Dressulhaus magnet in Eq.~(\ref{Dresseihaus_continuous_model}) at $\alpha_{R}=\alpha_{D}$ are connected by a unitary transformation that exchanges the Pauli matrices, provided the magnetization is rotated accordingly such that the exchange coupling $J_{ex}{\bf S}\cdot{\boldsymbol\sigma}$ is invariant and the energy spectrum remains the same 
\begin{eqnarray}
&&\left\{\sigma^{x},\sigma^{y},\sigma^{z}\right\}\rightarrow\left\{\sigma^{y},\sigma^{x},-\sigma^{z}\right\},
\nonumber \\
&&\left\{S^{x},S^{y},S^{z}\right\}\rightarrow\left\{S^{y},S^{x},-S^{z}\right\},
\label{Rashba_Dressulhaus_spin_mag_rotation}
\end{eqnarray}
The unitary transformation is defined and given by
\begin{eqnarray}
UH_{R}U^{\dag}=H_{D},\;\;\;U=\left(
\begin{array}{cc}
 & (1-i)/\sqrt{2} \\
(1+i)/\sqrt{2} & \\ 
\end{array}
\right).\;\;\;\;\;\;
\end{eqnarray}
The eigenstates $H_{R}|\psi_{R}\rangle=E|\psi_{R}\rangle$ and $H_{D}|\psi_{D}\rangle=E|\psi_{D}\rangle$ that have the same energy $E$ are transformed by $|\psi_{D}\rangle=U|\psi_{R}\rangle$. The current operator is transformed by
\begin{eqnarray}
J_{D}=\frac{1}{\hbar}\frac{\partial H_{D}}{\partial k_{x}}=U\frac{1}{\hbar}\frac{\partial H_{R}}{\partial k_{x}}U^{\dag}=UJ_{R}U^{\dag}.
\end{eqnarray}
Consequently, the expectation values of the charge current operator for the two systems are the same
\begin{eqnarray}
\langle J_{D}\rangle&=&\sum_{D}\langle\psi_{D}|J_{D}|\psi_{D}\rangle f(E)
\nonumber \\
&=&\sum_{R}\langle\psi_{R}|J_{R}|\psi_{R}\rangle f(E)=\langle J_{R}\rangle,
\end{eqnarray}
hence the two systems have the same spatial profile of edge current. The existence of edge spin currents can be argued in the same way, provided the spin quantization axis is rotated accordingly following Eqs.~(\ref{spin_current_definition_general}) and (\ref{Rashba_Dressulhaus_spin_mag_rotation}).

Combining the rotation of magnetization in Eq.~(\ref{Rashba_Dressulhaus_spin_mag_rotation}) with the results in Sec.~\ref{sec:Rashba_strip}, we conclude that a strip of Dresselhaus system going along ${\hat{\bf x}}$ direction manifests the following features: (i) ${\bf S=0}$: In the sbsence of magnetization, there exists bulk spin currents $J_{x}^{x}$ and $J_{y}^{y}$, and a helical edge spin current $J_{x}^{z}$. (ii) ${\bf S}\parallel{\hat{\bf x}}$: For the magnetization along the strip, the exists nonchiral edge current $J_{x}^{0}$ and a transverse spin polarization $\langle\sigma^{z}\rangle$. (iii) ${\bf S}\parallel{\hat{\bf y}}$: The in-plane magnetization perpendicular to the strip has no particular feature other than the spin currents stated above. (iv) ${\bf S}\parallel{\hat{\bf z}}$: The out-of-plane magnetization induces a chiral edge current $J_{x}^{0}$ and a transverse spin polarization $\langle\sigma^{x}\rangle$. Finally, the above argument also indicates that any other form of spin-orbit coupling that can be obtained from the RSOC through a unitary transformation will also have the laminar edge currents and transverse spin polarization, provided the magnetization orients at the appropriate direction.

\bibliography{Literatur}

\begin{thebibliography}{88}%
\makeatletter
\providecommand \@ifxundefined [1]{%
 \@ifx{#1\undefined}
}%
\providecommand \@ifnum [1]{%
 \ifnum #1\expandafter \@firstoftwo
 \else \expandafter \@secondoftwo
 \fi
}%
\providecommand \@ifx [1]{%
 \ifx #1\expandafter \@firstoftwo
 \else \expandafter \@secondoftwo
 \fi
}%
\providecommand \natexlab [1]{#1}%
\providecommand \enquote  [1]{``#1''}%
\providecommand \bibnamefont  [1]{#1}%
\providecommand \bibfnamefont [1]{#1}%
\providecommand \citenamefont [1]{#1}%
\providecommand \href@noop [0]{\@secondoftwo}%
\providecommand \href [0]{\begingroup \@sanitize@url \@href}%
\providecommand \@href[1]{\@@startlink{#1}\@@href}%
\providecommand \@@href[1]{\endgroup#1\@@endlink}%
\providecommand \@sanitize@url [0]{\catcode `\\12\catcode `\$12\catcode
  `\&12\catcode `\#12\catcode `\^12\catcode `\_12\catcode `\%12\relax}%
\providecommand \@@startlink[1]{}%
\providecommand \@@endlink[0]{}%
\providecommand \url  [0]{\begingroup\@sanitize@url \@url }%
\providecommand \@url [1]{\endgroup\@href {#1}{\urlprefix }}%
\providecommand \urlprefix  [0]{URL }%
\providecommand \Eprint [0]{\href }%
\providecommand \doibase [0]{http://dx.doi.org/}%
\providecommand \selectlanguage [0]{\@gobble}%
\providecommand \bibinfo  [0]{\@secondoftwo}%
\providecommand \bibfield  [0]{\@secondoftwo}%
\providecommand \translation [1]{[#1]}%
\providecommand \BibitemOpen [0]{}%
\providecommand \bibitemStop [0]{}%
\providecommand \bibitemNoStop [0]{.\EOS\space}%
\providecommand \EOS [0]{\spacefactor3000\relax}%
\providecommand \BibitemShut  [1]{\csname bibitem#1\endcsname}%
\let\auto@bib@innerbib\@empty
%</preamble>
\bibitem [{\citenamefont {Bychkov}\ and\ \citenamefont
  {Rashba}(1984{\natexlab{a}})}]{Bychkov84}%
  \BibitemOpen
  \bibfield  {author} {\bibinfo {author} {\bibfnamefont {Y.~A.}\ \bibnamefont
  {Bychkov}}\ and\ \bibinfo {author} {\bibfnamefont {E.~I.}\ \bibnamefont
  {Rashba}},\ }\href {\doibase 10.1103/PhysRevB.93.155104} {\bibfield
  {journal} {\bibinfo  {journal} {JETP Lett.}\ }\textbf {\bibinfo {volume}
  {39}},\ \bibinfo {pages} {78} (\bibinfo {year}
  {1984}{\natexlab{a}})}\BibitemShut {NoStop}%
\bibitem [{\citenamefont {Bychkov}\ and\ \citenamefont
  {Rashba}(1984{\natexlab{b}})}]{Bychkov84_2}%
  \BibitemOpen
  \bibfield  {author} {\bibinfo {author} {\bibfnamefont {Y.~A.}\ \bibnamefont
  {Bychkov}}\ and\ \bibinfo {author} {\bibfnamefont {E.~I.}\ \bibnamefont
  {Rashba}},\ }\href {\doibase 10.1088/0022-3719/17/33/015} {\bibfield
  {journal} {\bibinfo  {journal} {J. Phys. C: Solid State Phys.}\ }\textbf
  {\bibinfo {volume} {17}},\ \bibinfo {pages} {6039} (\bibinfo {year}
  {1984}{\natexlab{b}})}\BibitemShut {NoStop}%
\bibitem [{\citenamefont {Manchon}\ and\ \citenamefont
  {Zhang}(2008)}]{Manchon08}%
  \BibitemOpen
  \bibfield  {author} {\bibinfo {author} {\bibfnamefont {A.}~\bibnamefont
  {Manchon}}\ and\ \bibinfo {author} {\bibfnamefont {S.}~\bibnamefont
  {Zhang}},\ }\href {\doibase 10.1103/PhysRevB.78.212405} {\bibfield  {journal}
  {\bibinfo  {journal} {Phys. Rev. B}\ }\textbf {\bibinfo {volume} {78}},\
  \bibinfo {pages} {212405} (\bibinfo {year} {2008})}\BibitemShut {NoStop}%
\bibitem [{\citenamefont {Manchon}\ and\ \citenamefont
  {Zhang}(2009)}]{Manchon09}%
  \BibitemOpen
  \bibfield  {author} {\bibinfo {author} {\bibfnamefont {A.}~\bibnamefont
  {Manchon}}\ and\ \bibinfo {author} {\bibfnamefont {S.}~\bibnamefont
  {Zhang}},\ }\href {\doibase 10.1103/PhysRevB.79.094422} {\bibfield  {journal}
  {\bibinfo  {journal} {Phys. Rev. B}\ }\textbf {\bibinfo {volume} {79}},\
  \bibinfo {pages} {094422} (\bibinfo {year} {2009})}\BibitemShut {NoStop}%
\bibitem [{\citenamefont {Haney}\ and\ \citenamefont {Stiles}(2010)}]{Haney10}%
  \BibitemOpen
  \bibfield  {author} {\bibinfo {author} {\bibfnamefont {P.~M.}\ \bibnamefont
  {Haney}}\ and\ \bibinfo {author} {\bibfnamefont {M.~D.}\ \bibnamefont
  {Stiles}},\ }\href {\doibase 10.1103/PhysRevLett.105.126602} {\bibfield
  {journal} {\bibinfo  {journal} {Phys. Rev. Lett.}\ }\textbf {\bibinfo
  {volume} {105}},\ \bibinfo {pages} {126602} (\bibinfo {year}
  {2010})}\BibitemShut {NoStop}%
\bibitem [{\citenamefont {Gambardella}\ and\ \citenamefont
  {Miron}(2011)}]{Gambardella11}%
  \BibitemOpen
  \bibfield  {author} {\bibinfo {author} {\bibfnamefont {P.}~\bibnamefont
  {Gambardella}}\ and\ \bibinfo {author} {\bibfnamefont {I.~M.}\ \bibnamefont
  {Miron}},\ }\href {\doibase 10.1098/rsta.2010.0336} {\bibfield  {journal}
  {\bibinfo  {journal} {Phil. Trans. R. Soc. A}\ }\textbf {\bibinfo {volume}
  {369}},\ \bibinfo {pages} {3175} (\bibinfo {year} {2011})}\BibitemShut
  {NoStop}%
\bibitem [{\citenamefont {Pesin}\ and\ \citenamefont
  {MacDonald}(2012)}]{Pesin12}%
  \BibitemOpen
  \bibfield  {author} {\bibinfo {author} {\bibfnamefont {D.~A.}\ \bibnamefont
  {Pesin}}\ and\ \bibinfo {author} {\bibfnamefont {A.~H.}\ \bibnamefont
  {MacDonald}},\ }\href {\doibase 10.1103/PhysRevB.86.014416} {\bibfield
  {journal} {\bibinfo  {journal} {Phys. Rev. B}\ }\textbf {\bibinfo {volume}
  {86}},\ \bibinfo {pages} {014416} (\bibinfo {year} {2012})}\BibitemShut
  {NoStop}%
\bibitem [{\citenamefont {Haney}\ \emph {et~al.}(2013)\citenamefont {Haney},
  \citenamefont {Lee}, \citenamefont {Lee}, \citenamefont {Manchon},\ and\
  \citenamefont {Stiles}}]{Haney13}%
  \BibitemOpen
  \bibfield  {author} {\bibinfo {author} {\bibfnamefont {P.~M.}\ \bibnamefont
  {Haney}}, \bibinfo {author} {\bibfnamefont {H.-W.}\ \bibnamefont {Lee}},
  \bibinfo {author} {\bibfnamefont {K.-J.}\ \bibnamefont {Lee}}, \bibinfo
  {author} {\bibfnamefont {A.}~\bibnamefont {Manchon}}, \ and\ \bibinfo
  {author} {\bibfnamefont {M.~D.}\ \bibnamefont {Stiles}},\ }\href {\doibase
  10.1103/PhysRevB.88.214417} {\bibfield  {journal} {\bibinfo  {journal} {Phys.
  Rev. B}\ }\textbf {\bibinfo {volume} {88}},\ \bibinfo {pages} {214417}
  (\bibinfo {year} {2013})}\BibitemShut {NoStop}%
\bibitem [{\citenamefont {Hirsch}(1999)}]{Hirsch99}%
  \BibitemOpen
  \bibfield  {author} {\bibinfo {author} {\bibfnamefont {J.~E.}\ \bibnamefont
  {Hirsch}},\ }\href {\doibase 10.1103/PhysRevLett.83.1834} {\bibfield
  {journal} {\bibinfo  {journal} {Phys. Rev. Lett.}\ }\textbf {\bibinfo
  {volume} {83}},\ \bibinfo {pages} {1834} (\bibinfo {year}
  {1999})}\BibitemShut {NoStop}%
\bibitem [{\citenamefont {Sinova}\ \emph {et~al.}(2004)\citenamefont {Sinova},
  \citenamefont {Culcer}, \citenamefont {Niu}, \citenamefont {Sinitsyn},
  \citenamefont {Jungwirth},\ and\ \citenamefont {MacDonald}}]{Sinova04}%
  \BibitemOpen
  \bibfield  {author} {\bibinfo {author} {\bibfnamefont {J.}~\bibnamefont
  {Sinova}}, \bibinfo {author} {\bibfnamefont {D.}~\bibnamefont {Culcer}},
  \bibinfo {author} {\bibfnamefont {Q.}~\bibnamefont {Niu}}, \bibinfo {author}
  {\bibfnamefont {N.~A.}\ \bibnamefont {Sinitsyn}}, \bibinfo {author}
  {\bibfnamefont {T.}~\bibnamefont {Jungwirth}}, \ and\ \bibinfo {author}
  {\bibfnamefont {A.~H.}\ \bibnamefont {MacDonald}},\ }\href {\doibase
  10.1103/PhysRevLett.92.126603} {\bibfield  {journal} {\bibinfo  {journal}
  {Phys. Rev. Lett.}\ }\textbf {\bibinfo {volume} {92}},\ \bibinfo {pages}
  {126603} (\bibinfo {year} {2004})}\BibitemShut {NoStop}%
\bibitem [{\citenamefont {Kato}\ \emph {et~al.}(2004)\citenamefont {Kato},
  \citenamefont {Myers}, \citenamefont {Gossard},\ and\ \citenamefont
  {Awschalom}}]{Kato04}%
  \BibitemOpen
  \bibfield  {author} {\bibinfo {author} {\bibfnamefont {Y.~K.}\ \bibnamefont
  {Kato}}, \bibinfo {author} {\bibfnamefont {R.~C.}\ \bibnamefont {Myers}},
  \bibinfo {author} {\bibfnamefont {A.~C.}\ \bibnamefont {Gossard}}, \ and\
  \bibinfo {author} {\bibfnamefont {D.~D.}\ \bibnamefont {Awschalom}},\ }\href
  {\doibase 10.1126/science.1105514} {\bibfield  {journal} {\bibinfo  {journal}
  {Science}\ }\textbf {\bibinfo {volume} {306}},\ \bibinfo {pages} {1910}
  (\bibinfo {year} {2004})}\BibitemShut {NoStop}%
\bibitem [{\citenamefont {Valenzuela}\ and\ \citenamefont
  {Tinkham}(2006)}]{Valenzuela06}%
  \BibitemOpen
  \bibfield  {author} {\bibinfo {author} {\bibfnamefont {S.~O.}\ \bibnamefont
  {Valenzuela}}\ and\ \bibinfo {author} {\bibfnamefont {M.}~\bibnamefont
  {Tinkham}},\ }\href {\doibase 10.1038/nature04937} {\bibfield  {journal}
  {\bibinfo  {journal} {Nature}\ }\textbf {\bibinfo {volume} {442}},\ \bibinfo
  {pages} {176} (\bibinfo {year} {2006})}\BibitemShut {NoStop}%
\bibitem [{\citenamefont {Sinova}\ \emph {et~al.}(2015)\citenamefont {Sinova},
  \citenamefont {Valenzuela}, \citenamefont {Wunderlich}, \citenamefont
  {Back},\ and\ \citenamefont {Jungwirth}}]{Sinova15}%
  \BibitemOpen
  \bibfield  {author} {\bibinfo {author} {\bibfnamefont {J.}~\bibnamefont
  {Sinova}}, \bibinfo {author} {\bibfnamefont {S.~O.}\ \bibnamefont
  {Valenzuela}}, \bibinfo {author} {\bibfnamefont {J.}~\bibnamefont
  {Wunderlich}}, \bibinfo {author} {\bibfnamefont {C.~H.}\ \bibnamefont
  {Back}}, \ and\ \bibinfo {author} {\bibfnamefont {T.}~\bibnamefont
  {Jungwirth}},\ }\href {\doibase 10.1103/RevModPhys.87.1213} {\bibfield
  {journal} {\bibinfo  {journal} {Rev. Mod. Phys.}\ }\textbf {\bibinfo {volume}
  {87}},\ \bibinfo {pages} {1213} (\bibinfo {year} {2015})}\BibitemShut
  {NoStop}%
\bibitem [{\citenamefont {Usaj}\ and\ \citenamefont {Balseiro}(2005)}]{Usaj05}%
  \BibitemOpen
  \bibfield  {author} {\bibinfo {author} {\bibfnamefont {G.}~\bibnamefont
  {Usaj}}\ and\ \bibinfo {author} {\bibfnamefont {C.~A.}\ \bibnamefont
  {Balseiro}},\ }\href {\doibase 10.1209/epl/i2005-10266-0} {\bibfield
  {journal} {\bibinfo  {journal} {EPL}\ }\textbf {\bibinfo {volume} {72}},\
  \bibinfo {pages} {631} (\bibinfo {year} {2005})}\BibitemShut {NoStop}%
\bibitem [{\citenamefont {Reynoso}\ \emph {et~al.}(2004)\citenamefont
  {Reynoso}, \citenamefont {Usaj}, \citenamefont {S\'anchez},\ and\
  \citenamefont {Balseiro}}]{Reynoso04}%
  \BibitemOpen
  \bibfield  {author} {\bibinfo {author} {\bibfnamefont {A.}~\bibnamefont
  {Reynoso}}, \bibinfo {author} {\bibfnamefont {G.}~\bibnamefont {Usaj}},
  \bibinfo {author} {\bibfnamefont {M.~J.}\ \bibnamefont {S\'anchez}}, \ and\
  \bibinfo {author} {\bibfnamefont {C.~A.}\ \bibnamefont {Balseiro}},\ }\href
  {\doibase 10.1103/PhysRevB.70.235344} {\bibfield  {journal} {\bibinfo
  {journal} {Phys. Rev. B}\ }\textbf {\bibinfo {volume} {70}},\ \bibinfo
  {pages} {235344} (\bibinfo {year} {2004})}\BibitemShut {NoStop}%
\bibitem [{\citenamefont {Grigoryan}\ \emph {et~al.}(2009)\citenamefont
  {Grigoryan}, \citenamefont {Matos~Abiague},\ and\ \citenamefont
  {Badalyan}}]{Grigoryan09}%
  \BibitemOpen
  \bibfield  {author} {\bibinfo {author} {\bibfnamefont {V.~L.}\ \bibnamefont
  {Grigoryan}}, \bibinfo {author} {\bibfnamefont {A.}~\bibnamefont
  {Matos~Abiague}}, \ and\ \bibinfo {author} {\bibfnamefont {S.~M.}\
  \bibnamefont {Badalyan}},\ }\href {\doibase 10.1103/PhysRevB.80.165320}
  {\bibfield  {journal} {\bibinfo  {journal} {Phys. Rev. B}\ }\textbf {\bibinfo
  {volume} {80}},\ \bibinfo {pages} {165320} (\bibinfo {year}
  {2009})}\BibitemShut {NoStop}%
\bibitem [{\citenamefont {Sablikov}\ \emph {et~al.}(2008)\citenamefont
  {Sablikov}, \citenamefont {Sukhanov},\ and\ \citenamefont
  {Tkach}}]{Sablikov08}%
  \BibitemOpen
  \bibfield  {author} {\bibinfo {author} {\bibfnamefont {V.~A.}\ \bibnamefont
  {Sablikov}}, \bibinfo {author} {\bibfnamefont {A.~A.}\ \bibnamefont
  {Sukhanov}}, \ and\ \bibinfo {author} {\bibfnamefont {Y.~Y.}\ \bibnamefont
  {Tkach}},\ }\href {\doibase 10.1103/PhysRevB.78.153302} {\bibfield  {journal}
  {\bibinfo  {journal} {Phys. Rev. B}\ }\textbf {\bibinfo {volume} {78}},\
  \bibinfo {pages} {153302} (\bibinfo {year} {2008})}\BibitemShut {NoStop}%
\bibitem [{\citenamefont {de~Sousa}\ \emph {et~al.}(2021)\citenamefont
  {de~Sousa}, \citenamefont {Sigrist},\ and\ \citenamefont
  {Chen}}]{deSousa21_RSOC_nanoribbon}%
  \BibitemOpen
  \bibfield  {author} {\bibinfo {author} {\bibfnamefont {M.~S.~M.}\
  \bibnamefont {de~Sousa}}, \bibinfo {author} {\bibfnamefont {M.}~\bibnamefont
  {Sigrist}}, \ and\ \bibinfo {author} {\bibfnamefont {W.}~\bibnamefont
  {Chen}},\ }\href {\doibase 10.1103/PhysRevResearch.3.033021} {\bibfield
  {journal} {\bibinfo  {journal} {Phys. Rev. Research}\ }\textbf {\bibinfo
  {volume} {3}},\ \bibinfo {pages} {033021} (\bibinfo {year}
  {2021})}\BibitemShut {NoStop}%
\bibitem [{\citenamefont {Novoselov}\ \emph {et~al.}(2005)\citenamefont
  {Novoselov}, \citenamefont {Geim}, \citenamefont {Morozov}, \citenamefont
  {Jiang}, \citenamefont {Katsnelson}, \citenamefont {Grigorieva},
  \citenamefont {Dubonos},\ and\ \citenamefont {Firsov}}]{Novoselov05}%
  \BibitemOpen
  \bibfield  {author} {\bibinfo {author} {\bibfnamefont {K.~S.}\ \bibnamefont
  {Novoselov}}, \bibinfo {author} {\bibfnamefont {A.~K.}\ \bibnamefont {Geim}},
  \bibinfo {author} {\bibfnamefont {S.~V.}\ \bibnamefont {Morozov}}, \bibinfo
  {author} {\bibfnamefont {D.}~\bibnamefont {Jiang}}, \bibinfo {author}
  {\bibfnamefont {M.~I.}\ \bibnamefont {Katsnelson}}, \bibinfo {author}
  {\bibfnamefont {I.~V.}\ \bibnamefont {Grigorieva}}, \bibinfo {author}
  {\bibfnamefont {S.~V.}\ \bibnamefont {Dubonos}}, \ and\ \bibinfo {author}
  {\bibfnamefont {A.~A.}\ \bibnamefont {Firsov}},\ }\href {\doibase
  10.1038/nature04233} {\bibfield  {journal} {\bibinfo  {journal} {Nature}\
  }\textbf {\bibinfo {volume} {438}},\ \bibinfo {pages} {197} (\bibinfo {year}
  {2005})}\BibitemShut {NoStop}%
\bibitem [{\citenamefont {Zhou}\ \emph {et~al.}(2006)\citenamefont {Zhou},
  \citenamefont {Gweon}, \citenamefont {Graf}, \citenamefont {Fedorov},
  \citenamefont {Spataru}, \citenamefont {Diehl}, \citenamefont {Kopelevich},
  \citenamefont {Lee}, \citenamefont {Louie},\ and\ \citenamefont
  {Lanzara}}]{Zhou06}%
  \BibitemOpen
  \bibfield  {author} {\bibinfo {author} {\bibfnamefont {S.~Y.}\ \bibnamefont
  {Zhou}}, \bibinfo {author} {\bibfnamefont {G.-H.}\ \bibnamefont {Gweon}},
  \bibinfo {author} {\bibfnamefont {J.}~\bibnamefont {Graf}}, \bibinfo {author}
  {\bibfnamefont {A.~V.}\ \bibnamefont {Fedorov}}, \bibinfo {author}
  {\bibfnamefont {C.~D.}\ \bibnamefont {Spataru}}, \bibinfo {author}
  {\bibfnamefont {R.~D.}\ \bibnamefont {Diehl}}, \bibinfo {author}
  {\bibfnamefont {Y.}~\bibnamefont {Kopelevich}}, \bibinfo {author}
  {\bibfnamefont {D.-H.}\ \bibnamefont {Lee}}, \bibinfo {author} {\bibfnamefont
  {S.~G.}\ \bibnamefont {Louie}}, \ and\ \bibinfo {author} {\bibfnamefont
  {A.}~\bibnamefont {Lanzara}},\ }\href {\doibase 10.1038/nphys393} {\bibfield
  {journal} {\bibinfo  {journal} {Nat. Phys.}\ }\textbf {\bibinfo {volume}
  {2}},\ \bibinfo {pages} {595} (\bibinfo {year} {2006})}\BibitemShut {NoStop}%
\bibitem [{\citenamefont {Geim}\ and\ \citenamefont
  {Novoselov}(2007)}]{Geim07}%
  \BibitemOpen
  \bibfield  {author} {\bibinfo {author} {\bibfnamefont {A.~K.}\ \bibnamefont
  {Geim}}\ and\ \bibinfo {author} {\bibfnamefont {K.~S.}\ \bibnamefont
  {Novoselov}},\ }\href {\doibase 10.1038/nmat1849} {\bibfield  {journal}
  {\bibinfo  {journal} {Nat. Mater.}\ }\textbf {\bibinfo {volume} {6}},\
  \bibinfo {pages} {183} (\bibinfo {year} {2007})}\BibitemShut {NoStop}%
\bibitem [{\citenamefont {Castro~Neto}\ \emph {et~al.}(2009)\citenamefont
  {Castro~Neto}, \citenamefont {Guinea}, \citenamefont {Peres}, \citenamefont
  {Novoselov},\ and\ \citenamefont {Geim}}]{CastroNeto09}%
  \BibitemOpen
  \bibfield  {author} {\bibinfo {author} {\bibfnamefont {A.~H.}\ \bibnamefont
  {Castro~Neto}}, \bibinfo {author} {\bibfnamefont {F.}~\bibnamefont {Guinea}},
  \bibinfo {author} {\bibfnamefont {N.~M.~R.}\ \bibnamefont {Peres}}, \bibinfo
  {author} {\bibfnamefont {K.~S.}\ \bibnamefont {Novoselov}}, \ and\ \bibinfo
  {author} {\bibfnamefont {A.~K.}\ \bibnamefont {Geim}},\ }\href {\doibase
  10.1103/RevModPhys.81.109} {\bibfield  {journal} {\bibinfo  {journal} {Rev.
  Mod. Phys.}\ }\textbf {\bibinfo {volume} {81}},\ \bibinfo {pages} {109}
  (\bibinfo {year} {2009})}\BibitemShut {NoStop}%
\bibitem [{\citenamefont {Geim}(2009)}]{Geim09}%
  \BibitemOpen
  \bibfield  {author} {\bibinfo {author} {\bibfnamefont {A.~K.}\ \bibnamefont
  {Geim}},\ }\href {\doibase 10.1126/science.1158877} {\bibfield  {journal}
  {\bibinfo  {journal} {Science}\ }\textbf {\bibinfo {volume} {324}},\ \bibinfo
  {pages} {1530} (\bibinfo {year} {2009})}\BibitemShut {NoStop}%
\bibitem [{\citenamefont {Das~Sarma}\ \emph {et~al.}(2011)\citenamefont
  {Das~Sarma}, \citenamefont {Adam}, \citenamefont {Hwang},\ and\ \citenamefont
  {Rossi}}]{DasSarma11}%
  \BibitemOpen
  \bibfield  {author} {\bibinfo {author} {\bibfnamefont {S.}~\bibnamefont
  {Das~Sarma}}, \bibinfo {author} {\bibfnamefont {S.}~\bibnamefont {Adam}},
  \bibinfo {author} {\bibfnamefont {E.~H.}\ \bibnamefont {Hwang}}, \ and\
  \bibinfo {author} {\bibfnamefont {E.}~\bibnamefont {Rossi}},\ }\href
  {\doibase 10.1103/RevModPhys.83.407} {\bibfield  {journal} {\bibinfo
  {journal} {Rev. Mod. Phys.}\ }\textbf {\bibinfo {volume} {83}},\ \bibinfo
  {pages} {407} (\bibinfo {year} {2011})}\BibitemShut {NoStop}%
\bibitem [{\citenamefont {Fujita}\ \emph {et~al.}(1996)\citenamefont {Fujita},
  \citenamefont {Wakabayashi}, \citenamefont {Nakada},\ and\ \citenamefont
  {Kusakabe}}]{Fujita96}%
  \BibitemOpen
  \bibfield  {author} {\bibinfo {author} {\bibfnamefont {M.}~\bibnamefont
  {Fujita}}, \bibinfo {author} {\bibfnamefont {K.}~\bibnamefont {Wakabayashi}},
  \bibinfo {author} {\bibfnamefont {K.}~\bibnamefont {Nakada}}, \ and\ \bibinfo
  {author} {\bibfnamefont {K.}~\bibnamefont {Kusakabe}},\ }\href {\doibase
  10.1143/JPSJ.65.1920} {\bibfield  {journal} {\bibinfo  {journal} {Journal of
  the Physical Society of Japan}\ }\textbf {\bibinfo {volume} {65}},\ \bibinfo
  {pages} {1920} (\bibinfo {year} {1996})}\BibitemShut {NoStop}%
\bibitem [{\citenamefont {Nakada}\ \emph {et~al.}(1996)\citenamefont {Nakada},
  \citenamefont {Fujita}, \citenamefont {Dresselhaus},\ and\ \citenamefont
  {Dresselhaus}}]{Nakada96}%
  \BibitemOpen
  \bibfield  {author} {\bibinfo {author} {\bibfnamefont {K.}~\bibnamefont
  {Nakada}}, \bibinfo {author} {\bibfnamefont {M.}~\bibnamefont {Fujita}},
  \bibinfo {author} {\bibfnamefont {G.}~\bibnamefont {Dresselhaus}}, \ and\
  \bibinfo {author} {\bibfnamefont {M.~S.}\ \bibnamefont {Dresselhaus}},\
  }\href {\doibase 10.1103/PhysRevB.54.17954} {\bibfield  {journal} {\bibinfo
  {journal} {Phys. Rev. B}\ }\textbf {\bibinfo {volume} {54}},\ \bibinfo
  {pages} {17954} (\bibinfo {year} {1996})}\BibitemShut {NoStop}%
\bibitem [{\citenamefont {Brey}\ and\ \citenamefont {Fertig}(2006)}]{Brey06}%
  \BibitemOpen
  \bibfield  {author} {\bibinfo {author} {\bibfnamefont {L.}~\bibnamefont
  {Brey}}\ and\ \bibinfo {author} {\bibfnamefont {H.~A.}\ \bibnamefont
  {Fertig}},\ }\href {\doibase 10.1103/PhysRevB.73.235411} {\bibfield
  {journal} {\bibinfo  {journal} {Phys. Rev. B}\ }\textbf {\bibinfo {volume}
  {73}},\ \bibinfo {pages} {235411} (\bibinfo {year} {2006})}\BibitemShut
  {NoStop}%
\bibitem [{\citenamefont {Ezawa}(2006)}]{Ezawa06}%
  \BibitemOpen
  \bibfield  {author} {\bibinfo {author} {\bibfnamefont {M.}~\bibnamefont
  {Ezawa}},\ }\href {\doibase 10.1103/PhysRevB.73.045432} {\bibfield  {journal}
  {\bibinfo  {journal} {Phys. Rev. B}\ }\textbf {\bibinfo {volume} {73}},\
  \bibinfo {pages} {045432} (\bibinfo {year} {2006})}\BibitemShut {NoStop}%
\bibitem [{\citenamefont {Peres}\ \emph {et~al.}(2006)\citenamefont {Peres},
  \citenamefont {Guinea},\ and\ \citenamefont {Castro~Neto}}]{Peres06}%
  \BibitemOpen
  \bibfield  {author} {\bibinfo {author} {\bibfnamefont {N.~M.~R.}\
  \bibnamefont {Peres}}, \bibinfo {author} {\bibfnamefont {F.}~\bibnamefont
  {Guinea}}, \ and\ \bibinfo {author} {\bibfnamefont {A.~H.}\ \bibnamefont
  {Castro~Neto}},\ }\href {\doibase 10.1103/PhysRevB.73.125411} {\bibfield
  {journal} {\bibinfo  {journal} {Phys. Rev. B}\ }\textbf {\bibinfo {volume}
  {73}},\ \bibinfo {pages} {125411} (\bibinfo {year} {2006})}\BibitemShut
  {NoStop}%
\bibitem [{\citenamefont {Akhmerov}\ and\ \citenamefont
  {Beenakker}(2008)}]{Akhmerov08}%
  \BibitemOpen
  \bibfield  {author} {\bibinfo {author} {\bibfnamefont {A.~R.}\ \bibnamefont
  {Akhmerov}}\ and\ \bibinfo {author} {\bibfnamefont {C.~W.~J.}\ \bibnamefont
  {Beenakker}},\ }\href {\doibase 10.1103/PhysRevB.77.085423} {\bibfield
  {journal} {\bibinfo  {journal} {Phys. Rev. B}\ }\textbf {\bibinfo {volume}
  {77}},\ \bibinfo {pages} {085423} (\bibinfo {year} {2008})}\BibitemShut
  {NoStop}%
\bibitem [{\citenamefont {Tao}\ \emph {et~al.}(2011)\citenamefont {Tao},
  \citenamefont {Jiao}, \citenamefont {Yazyev}, \citenamefont {Chen},
  \citenamefont {Feng}, \citenamefont {Zhang}, \citenamefont {Capaz},
  \citenamefont {Tour}, \citenamefont {Zettl}, \citenamefont {Louie},
  \citenamefont {Dai},\ and\ \citenamefont {Crommie}}]{Tao11}%
  \BibitemOpen
  \bibfield  {author} {\bibinfo {author} {\bibfnamefont {C.}~\bibnamefont
  {Tao}}, \bibinfo {author} {\bibfnamefont {L.}~\bibnamefont {Jiao}}, \bibinfo
  {author} {\bibfnamefont {O.~V.}\ \bibnamefont {Yazyev}}, \bibinfo {author}
  {\bibfnamefont {Y.-C.}\ \bibnamefont {Chen}}, \bibinfo {author}
  {\bibfnamefont {J.}~\bibnamefont {Feng}}, \bibinfo {author} {\bibfnamefont
  {X.}~\bibnamefont {Zhang}}, \bibinfo {author} {\bibfnamefont {R.~B.}\
  \bibnamefont {Capaz}}, \bibinfo {author} {\bibfnamefont {J.~M.}\ \bibnamefont
  {Tour}}, \bibinfo {author} {\bibfnamefont {A.}~\bibnamefont {Zettl}},
  \bibinfo {author} {\bibfnamefont {S.~G.}\ \bibnamefont {Louie}}, \bibinfo
  {author} {\bibfnamefont {H.}~\bibnamefont {Dai}}, \ and\ \bibinfo {author}
  {\bibfnamefont {M.~F.}\ \bibnamefont {Crommie}},\ }\href {\doibase
  10.1038/nphys1991} {\bibfield  {journal} {\bibinfo  {journal} {Nat. Phys.}\
  }\textbf {\bibinfo {volume} {7}},\ \bibinfo {pages} {616} (\bibinfo {year}
  {2011})}\BibitemShut {NoStop}%
\bibitem [{\citenamefont {Ruffieux}\ \emph {et~al.}(2011)\citenamefont
  {Ruffieux}, \citenamefont {Wang}, \citenamefont {Yang}, \citenamefont
  {S\'{a}nchez-S\'{a}nchez}, \citenamefont {Liu}, \citenamefont {Dienel},
  \citenamefont {Talirz}, \citenamefont {Shinde}, \citenamefont {Pignedoli},
  \citenamefont {Passerone}, \citenamefont {Dumslaff}, \citenamefont {Feng},
  \citenamefont {M\"{u}llen},\ and\ \citenamefont {Fasel}}]{Ruffieux16}%
  \BibitemOpen
  \bibfield  {author} {\bibinfo {author} {\bibfnamefont {P.}~\bibnamefont
  {Ruffieux}}, \bibinfo {author} {\bibfnamefont {S.}~\bibnamefont {Wang}},
  \bibinfo {author} {\bibfnamefont {B.}~\bibnamefont {Yang}}, \bibinfo {author}
  {\bibfnamefont {C.}~\bibnamefont {S\'{a}nchez-S\'{a}nchez}}, \bibinfo
  {author} {\bibfnamefont {J.}~\bibnamefont {Liu}}, \bibinfo {author}
  {\bibfnamefont {T.}~\bibnamefont {Dienel}}, \bibinfo {author} {\bibfnamefont
  {L.}~\bibnamefont {Talirz}}, \bibinfo {author} {\bibfnamefont
  {P.}~\bibnamefont {Shinde}}, \bibinfo {author} {\bibfnamefont {C.~A.}\
  \bibnamefont {Pignedoli}}, \bibinfo {author} {\bibfnamefont {D.}~\bibnamefont
  {Passerone}}, \bibinfo {author} {\bibfnamefont {T.}~\bibnamefont {Dumslaff}},
  \bibinfo {author} {\bibfnamefont {X.}~\bibnamefont {Feng}}, \bibinfo {author}
  {\bibfnamefont {K.}~\bibnamefont {M\"{u}llen}}, \ and\ \bibinfo {author}
  {\bibfnamefont {R.}~\bibnamefont {Fasel}},\ }\href {\doibase
  10.1038/nature17151} {\bibfield  {journal} {\bibinfo  {journal} {Nature}\
  }\textbf {\bibinfo {volume} {531}},\ \bibinfo {pages} {489} (\bibinfo {year}
  {2011})}\BibitemShut {NoStop}%
\bibitem [{\citenamefont {Wakabayashi}\ \emph {et~al.}(1999)\citenamefont
  {Wakabayashi}, \citenamefont {Fujita}, \citenamefont {Ajiki},\ and\
  \citenamefont {Sigrist}}]{Wakabayashi99}%
  \BibitemOpen
  \bibfield  {author} {\bibinfo {author} {\bibfnamefont {K.}~\bibnamefont
  {Wakabayashi}}, \bibinfo {author} {\bibfnamefont {M.}~\bibnamefont {Fujita}},
  \bibinfo {author} {\bibfnamefont {H.}~\bibnamefont {Ajiki}}, \ and\ \bibinfo
  {author} {\bibfnamefont {M.}~\bibnamefont {Sigrist}},\ }\href {\doibase
  10.1103/PhysRevB.59.8271} {\bibfield  {journal} {\bibinfo  {journal} {Phys.
  Rev. B}\ }\textbf {\bibinfo {volume} {59}},\ \bibinfo {pages} {8271}
  (\bibinfo {year} {1999})}\BibitemShut {NoStop}%
\bibitem [{\citenamefont {Lee}\ \emph {et~al.}(2005)\citenamefont {Lee},
  \citenamefont {Son}, \citenamefont {Park}, \citenamefont {Han},\ and\
  \citenamefont {Yu}}]{Lee05}%
  \BibitemOpen
  \bibfield  {author} {\bibinfo {author} {\bibfnamefont {H.}~\bibnamefont
  {Lee}}, \bibinfo {author} {\bibfnamefont {Y.-W.}\ \bibnamefont {Son}},
  \bibinfo {author} {\bibfnamefont {N.}~\bibnamefont {Park}}, \bibinfo {author}
  {\bibfnamefont {S.}~\bibnamefont {Han}}, \ and\ \bibinfo {author}
  {\bibfnamefont {J.}~\bibnamefont {Yu}},\ }\href {\doibase
  10.1103/PhysRevB.72.174431} {\bibfield  {journal} {\bibinfo  {journal} {Phys.
  Rev. B}\ }\textbf {\bibinfo {volume} {72}},\ \bibinfo {pages} {174431}
  (\bibinfo {year} {2005})}\BibitemShut {NoStop}%
\bibitem [{\citenamefont {Pisani}\ \emph {et~al.}(2007)\citenamefont {Pisani},
  \citenamefont {Chan}, \citenamefont {Montanari},\ and\ \citenamefont
  {Harrison}}]{Pisani07}%
  \BibitemOpen
  \bibfield  {author} {\bibinfo {author} {\bibfnamefont {L.}~\bibnamefont
  {Pisani}}, \bibinfo {author} {\bibfnamefont {J.~A.}\ \bibnamefont {Chan}},
  \bibinfo {author} {\bibfnamefont {B.}~\bibnamefont {Montanari}}, \ and\
  \bibinfo {author} {\bibfnamefont {N.~M.}\ \bibnamefont {Harrison}},\ }\href
  {\doibase 10.1103/PhysRevB.75.064418} {\bibfield  {journal} {\bibinfo
  {journal} {Phys. Rev. B}\ }\textbf {\bibinfo {volume} {75}},\ \bibinfo
  {pages} {064418} (\bibinfo {year} {2007})}\BibitemShut {NoStop}%
\bibitem [{\citenamefont {Yang}\ \emph {et~al.}(2016)\citenamefont {Yang},
  \citenamefont {Tu}, \citenamefont {Kim}, \citenamefont {Wu}, \citenamefont
  {Wang}, \citenamefont {Alicea}, \citenamefont {Wu}, \citenamefont
  {Bockrath},\ and\ \citenamefont {Shi}}]{Yang16}%
  \BibitemOpen
  \bibfield  {author} {\bibinfo {author} {\bibfnamefont {B.}~\bibnamefont
  {Yang}}, \bibinfo {author} {\bibfnamefont {M.-F.}\ \bibnamefont {Tu}},
  \bibinfo {author} {\bibfnamefont {J.}~\bibnamefont {Kim}}, \bibinfo {author}
  {\bibfnamefont {Y.}~\bibnamefont {Wu}}, \bibinfo {author} {\bibfnamefont
  {H.}~\bibnamefont {Wang}}, \bibinfo {author} {\bibfnamefont {J.}~\bibnamefont
  {Alicea}}, \bibinfo {author} {\bibfnamefont {R.}~\bibnamefont {Wu}}, \bibinfo
  {author} {\bibfnamefont {M.}~\bibnamefont {Bockrath}}, \ and\ \bibinfo
  {author} {\bibfnamefont {J.}~\bibnamefont {Shi}},\ }\href {\doibase
  10.1088/2053-1583/3/3/031012} {\bibfield  {journal} {\bibinfo  {journal} {2D
  Mater.}\ }\textbf {\bibinfo {volume} {3}},\ \bibinfo {pages} {031012}
  (\bibinfo {year} {2016})}\BibitemShut {NoStop}%
\bibitem [{\citenamefont {Wang}\ \emph {et~al.}(2016)\citenamefont {Wang},
  \citenamefont {Ki}, \citenamefont {Khoo}, \citenamefont {Mauro},
  \citenamefont {Berger}, \citenamefont {Levitov},\ and\ \citenamefont
  {Morpurgo}}]{Wang16_3}%
  \BibitemOpen
  \bibfield  {author} {\bibinfo {author} {\bibfnamefont {Z.}~\bibnamefont
  {Wang}}, \bibinfo {author} {\bibfnamefont {D.-K.}\ \bibnamefont {Ki}},
  \bibinfo {author} {\bibfnamefont {J.~Y.}\ \bibnamefont {Khoo}}, \bibinfo
  {author} {\bibfnamefont {D.}~\bibnamefont {Mauro}}, \bibinfo {author}
  {\bibfnamefont {H.}~\bibnamefont {Berger}}, \bibinfo {author} {\bibfnamefont
  {L.~S.}\ \bibnamefont {Levitov}}, \ and\ \bibinfo {author} {\bibfnamefont
  {A.~F.}\ \bibnamefont {Morpurgo}},\ }\href {\doibase
  10.1103/PhysRevX.6.041020} {\bibfield  {journal} {\bibinfo  {journal} {Phys.
  Rev. X}\ }\textbf {\bibinfo {volume} {6}},\ \bibinfo {pages} {041020}
  (\bibinfo {year} {2016})}\BibitemShut {NoStop}%
\bibitem [{\citenamefont {Yang}\ \emph {et~al.}(2017)\citenamefont {Yang},
  \citenamefont {Lohmann}, \citenamefont {Barroso}, \citenamefont {Liao},
  \citenamefont {Lin}, \citenamefont {Liu}, \citenamefont {Bartels},
  \citenamefont {Watanabe}, \citenamefont {Taniguchi},\ and\ \citenamefont
  {Shi}}]{Yang17}%
  \BibitemOpen
  \bibfield  {author} {\bibinfo {author} {\bibfnamefont {B.}~\bibnamefont
  {Yang}}, \bibinfo {author} {\bibfnamefont {M.}~\bibnamefont {Lohmann}},
  \bibinfo {author} {\bibfnamefont {D.}~\bibnamefont {Barroso}}, \bibinfo
  {author} {\bibfnamefont {I.}~\bibnamefont {Liao}}, \bibinfo {author}
  {\bibfnamefont {Z.}~\bibnamefont {Lin}}, \bibinfo {author} {\bibfnamefont
  {Y.}~\bibnamefont {Liu}}, \bibinfo {author} {\bibfnamefont {L.}~\bibnamefont
  {Bartels}}, \bibinfo {author} {\bibfnamefont {K.}~\bibnamefont {Watanabe}},
  \bibinfo {author} {\bibfnamefont {T.}~\bibnamefont {Taniguchi}}, \ and\
  \bibinfo {author} {\bibfnamefont {J.}~\bibnamefont {Shi}},\ }\href {\doibase
  10.1103/PhysRevB.96.041409} {\bibfield  {journal} {\bibinfo  {journal} {Phys.
  Rev. B}\ }\textbf {\bibinfo {volume} {96}},\ \bibinfo {pages} {041409}
  (\bibinfo {year} {2017})}\BibitemShut {NoStop}%
\bibitem [{\citenamefont {Safeer}\ \emph {et~al.}(2019)\citenamefont {Safeer},
  \citenamefont {Ingla-Ayn\'{e}s}, \citenamefont {Herling}, \citenamefont
  {Garcia}, \citenamefont {Vila}, \citenamefont {Ontoso}, \citenamefont
  {Calvo}, \citenamefont {Roche},\ and\ \citenamefont {Hueso}}]{Safeer19}%
  \BibitemOpen
  \bibfield  {author} {\bibinfo {author} {\bibfnamefont {C.~K.}\ \bibnamefont
  {Safeer}}, \bibinfo {author} {\bibfnamefont {J.}~\bibnamefont
  {Ingla-Ayn\'{e}s}}, \bibinfo {author} {\bibfnamefont {F.}~\bibnamefont
  {Herling}}, \bibinfo {author} {\bibfnamefont {J.~H.}\ \bibnamefont {Garcia}},
  \bibinfo {author} {\bibfnamefont {M.}~\bibnamefont {Vila}}, \bibinfo {author}
  {\bibfnamefont {N.}~\bibnamefont {Ontoso}}, \bibinfo {author} {\bibfnamefont
  {M.~R.}\ \bibnamefont {Calvo}}, \bibinfo {author} {\bibfnamefont
  {S.}~\bibnamefont {Roche}}, \ and\ \bibinfo {author} {\bibfnamefont
  {F.}~\bibnamefont {Hueso}, \bibfnamefont {Luis E.and~Casanova}},\ }\href
  {\doibase 10.1021/acs.nanolett.8b04368} {\bibfield  {journal} {\bibinfo
  {journal} {Nano Lett.}\ }\textbf {\bibinfo {volume} {19}},\ \bibinfo {pages}
  {1074} (\bibinfo {year} {2019})}\BibitemShut {NoStop}%
\bibitem [{\citenamefont {Ghiasi}\ \emph {et~al.}(2019)\citenamefont {Ghiasi},
  \citenamefont {Kaverzin}, \citenamefont {Blah},\ and\ \citenamefont {van
  Wees}}]{Ghiasi19}%
  \BibitemOpen
  \bibfield  {author} {\bibinfo {author} {\bibfnamefont {T.~S.}\ \bibnamefont
  {Ghiasi}}, \bibinfo {author} {\bibfnamefont {A.~A.}\ \bibnamefont
  {Kaverzin}}, \bibinfo {author} {\bibfnamefont {P.~J.}\ \bibnamefont {Blah}},
  \ and\ \bibinfo {author} {\bibfnamefont {B.~J.}\ \bibnamefont {van Wees}},\
  }\href {\doibase 10.1021/acs.nanolett.9b01611} {\bibfield  {journal}
  {\bibinfo  {journal} {Nano Lett.}\ }\textbf {\bibinfo {volume} {19}},\
  \bibinfo {pages} {5959} (\bibinfo {year} {2019})}\BibitemShut {NoStop}%
\bibitem [{\citenamefont {Ben\'{i}tez}\ \emph {et~al.}(2020)\citenamefont
  {Ben\'{i}tez}, \citenamefont {Savero~Torres}, \citenamefont {Sierra},
  \citenamefont {Timmermans}, \citenamefont {Garcia}, \citenamefont {Roche},
  \citenamefont {Costache},\ and\ \citenamefont {Valenzuela}}]{Benitez20}%
  \BibitemOpen
  \bibfield  {author} {\bibinfo {author} {\bibfnamefont {L.~A.}\ \bibnamefont
  {Ben\'{i}tez}}, \bibinfo {author} {\bibfnamefont {W.}~\bibnamefont
  {Savero~Torres}}, \bibinfo {author} {\bibfnamefont {J.~F.}\ \bibnamefont
  {Sierra}}, \bibinfo {author} {\bibfnamefont {M.}~\bibnamefont {Timmermans}},
  \bibinfo {author} {\bibfnamefont {J.~H.}\ \bibnamefont {Garcia}}, \bibinfo
  {author} {\bibfnamefont {S.}~\bibnamefont {Roche}}, \bibinfo {author}
  {\bibfnamefont {M.~V.}\ \bibnamefont {Costache}}, \ and\ \bibinfo {author}
  {\bibfnamefont {S.~O.}\ \bibnamefont {Valenzuela}},\ }\href {\doibase
  10.1038/s41563-019-0575-1} {\bibfield  {journal} {\bibinfo  {journal} {Nat.
  Mater.}\ }\textbf {\bibinfo {volume} {19}},\ \bibinfo {pages} {170} (\bibinfo
  {year} {2020})}\BibitemShut {NoStop}%
\bibitem [{\citenamefont {Mendes}\ \emph {et~al.}(2015)\citenamefont {Mendes},
  \citenamefont {Alves~Santos}, \citenamefont {Meireles}, \citenamefont
  {Lacerda}, \citenamefont {Vilela-Le\~ao}, \citenamefont {Machado},
  \citenamefont {Rodr\'{\i}guez-Su\'arez}, \citenamefont {Azevedo},\ and\
  \citenamefont {Rezende}}]{Mendes15}%
  \BibitemOpen
  \bibfield  {author} {\bibinfo {author} {\bibfnamefont {J.~B.~S.}\
  \bibnamefont {Mendes}}, \bibinfo {author} {\bibfnamefont {O.}~\bibnamefont
  {Alves~Santos}}, \bibinfo {author} {\bibfnamefont {L.~M.}\ \bibnamefont
  {Meireles}}, \bibinfo {author} {\bibfnamefont {R.~G.}\ \bibnamefont
  {Lacerda}}, \bibinfo {author} {\bibfnamefont {L.~H.}\ \bibnamefont
  {Vilela-Le\~ao}}, \bibinfo {author} {\bibfnamefont {F.~L.~A.}\ \bibnamefont
  {Machado}}, \bibinfo {author} {\bibfnamefont {R.~L.}\ \bibnamefont
  {Rodr\'{\i}guez-Su\'arez}}, \bibinfo {author} {\bibfnamefont
  {A.}~\bibnamefont {Azevedo}}, \ and\ \bibinfo {author} {\bibfnamefont
  {S.~M.}\ \bibnamefont {Rezende}},\ }\href {\doibase
  10.1103/PhysRevLett.115.226601} {\bibfield  {journal} {\bibinfo  {journal}
  {Phys. Rev. Lett.}\ }\textbf {\bibinfo {volume} {115}},\ \bibinfo {pages}
  {226601} (\bibinfo {year} {2015})}\BibitemShut {NoStop}%
\bibitem [{\citenamefont {Wang}\ \emph {et~al.}(2015)\citenamefont {Wang},
  \citenamefont {Tang}, \citenamefont {Sachs}, \citenamefont {Barlas},\ and\
  \citenamefont {Shi}}]{Wang15_3}%
  \BibitemOpen
  \bibfield  {author} {\bibinfo {author} {\bibfnamefont {Z.}~\bibnamefont
  {Wang}}, \bibinfo {author} {\bibfnamefont {C.}~\bibnamefont {Tang}}, \bibinfo
  {author} {\bibfnamefont {R.}~\bibnamefont {Sachs}}, \bibinfo {author}
  {\bibfnamefont {Y.}~\bibnamefont {Barlas}}, \ and\ \bibinfo {author}
  {\bibfnamefont {J.}~\bibnamefont {Shi}},\ }\href {\doibase
  10.1103/PhysRevLett.114.016603} {\bibfield  {journal} {\bibinfo  {journal}
  {Phys. Rev. Lett.}\ }\textbf {\bibinfo {volume} {114}},\ \bibinfo {pages}
  {016603} (\bibinfo {year} {2015})}\BibitemShut {NoStop}%
\bibitem [{\citenamefont {Dushenko}\ \emph {et~al.}(2016)\citenamefont
  {Dushenko}, \citenamefont {Ago}, \citenamefont {Kawahara}, \citenamefont
  {Tsuda}, \citenamefont {Kuwabata}, \citenamefont {Takenobu}, \citenamefont
  {Shinjo}, \citenamefont {Ando},\ and\ \citenamefont
  {Shiraishi}}]{Dushenko16}%
  \BibitemOpen
  \bibfield  {author} {\bibinfo {author} {\bibfnamefont {S.}~\bibnamefont
  {Dushenko}}, \bibinfo {author} {\bibfnamefont {H.}~\bibnamefont {Ago}},
  \bibinfo {author} {\bibfnamefont {K.}~\bibnamefont {Kawahara}}, \bibinfo
  {author} {\bibfnamefont {T.}~\bibnamefont {Tsuda}}, \bibinfo {author}
  {\bibfnamefont {S.}~\bibnamefont {Kuwabata}}, \bibinfo {author}
  {\bibfnamefont {T.}~\bibnamefont {Takenobu}}, \bibinfo {author}
  {\bibfnamefont {T.}~\bibnamefont {Shinjo}}, \bibinfo {author} {\bibfnamefont
  {Y.}~\bibnamefont {Ando}}, \ and\ \bibinfo {author} {\bibfnamefont
  {M.}~\bibnamefont {Shiraishi}},\ }\href {\doibase
  10.1103/PhysRevLett.116.166102} {\bibfield  {journal} {\bibinfo  {journal}
  {Phys. Rev. Lett.}\ }\textbf {\bibinfo {volume} {116}},\ \bibinfo {pages}
  {166102} (\bibinfo {year} {2016})}\BibitemShut {NoStop}%
\bibitem [{\citenamefont {Leutenantsmeyer}\ \emph {et~al.}(2016)\citenamefont
  {Leutenantsmeyer}, \citenamefont {Kaverzin}, \citenamefont {Wojtaszek},\ and\
  \citenamefont {van Wees}}]{Leutenantsmeyer17}%
  \BibitemOpen
  \bibfield  {author} {\bibinfo {author} {\bibfnamefont {J.~C.}\ \bibnamefont
  {Leutenantsmeyer}}, \bibinfo {author} {\bibfnamefont {A.~A.}\ \bibnamefont
  {Kaverzin}}, \bibinfo {author} {\bibfnamefont {M.}~\bibnamefont {Wojtaszek}},
  \ and\ \bibinfo {author} {\bibfnamefont {B.~J.}\ \bibnamefont {van Wees}},\
  }\href {\doibase 10.1088/2053-1583/4/1/014001} {\bibfield  {journal}
  {\bibinfo  {journal} {2D Mater.}\ }\textbf {\bibinfo {volume} {4}},\ \bibinfo
  {pages} {014001} (\bibinfo {year} {2016})}\BibitemShut {NoStop}%
\bibitem [{\citenamefont {Rybkin}\ \emph {et~al.}(2018)\citenamefont {Rybkin},
  \citenamefont {Rybkina}, \citenamefont {Otrokov}, \citenamefont {Vilkov},
  \citenamefont {Klimovskikh}, \citenamefont {Petukhov}, \citenamefont
  {Filianina}, \citenamefont {Voroshnin}, \citenamefont {Rusinov},
  \citenamefont {Ernst}, \citenamefont {Arnau}, \citenamefont {Chulkov},\ and\
  \citenamefont {Shikin}}]{Rybkin18}%
  \BibitemOpen
  \bibfield  {author} {\bibinfo {author} {\bibfnamefont {A.~G.}\ \bibnamefont
  {Rybkin}}, \bibinfo {author} {\bibfnamefont {A.~A.}\ \bibnamefont {Rybkina}},
  \bibinfo {author} {\bibfnamefont {M.~M.}\ \bibnamefont {Otrokov}}, \bibinfo
  {author} {\bibfnamefont {O.~Y.}\ \bibnamefont {Vilkov}}, \bibinfo {author}
  {\bibfnamefont {I.~I.}\ \bibnamefont {Klimovskikh}}, \bibinfo {author}
  {\bibfnamefont {A.~E.}\ \bibnamefont {Petukhov}}, \bibinfo {author}
  {\bibfnamefont {M.~V.}\ \bibnamefont {Filianina}}, \bibinfo {author}
  {\bibfnamefont {V.~Y.}\ \bibnamefont {Voroshnin}}, \bibinfo {author}
  {\bibfnamefont {I.~P.}\ \bibnamefont {Rusinov}}, \bibinfo {author}
  {\bibfnamefont {A.}~\bibnamefont {Ernst}}, \bibinfo {author} {\bibfnamefont
  {A.}~\bibnamefont {Arnau}}, \bibinfo {author} {\bibfnamefont {E.~V.}\
  \bibnamefont {Chulkov}}, \ and\ \bibinfo {author} {\bibfnamefont {A.~M.}\
  \bibnamefont {Shikin}},\ }\href {\doibase 10.1021/acs.nanolett.7b01548}
  {\bibfield  {journal} {\bibinfo  {journal} {Nano Lett.}\ }\textbf {\bibinfo
  {volume} {18}},\ \bibinfo {pages} {1564} (\bibinfo {year}
  {2018})}\BibitemShut {NoStop}%
\bibitem [{\citenamefont {Liechtenstein}\ \emph {et~al.}(1987)\citenamefont
  {Liechtenstein}, \citenamefont {Katsnelson}, \citenamefont {Antropov},\ and\
  \citenamefont {Gubanov}}]{Liechtenstein87}%
  \BibitemOpen
  \bibfield  {author} {\bibinfo {author} {\bibfnamefont {A.}~\bibnamefont
  {Liechtenstein}}, \bibinfo {author} {\bibfnamefont {M.}~\bibnamefont
  {Katsnelson}}, \bibinfo {author} {\bibfnamefont {V.}~\bibnamefont
  {Antropov}}, \ and\ \bibinfo {author} {\bibfnamefont {V.}~\bibnamefont
  {Gubanov}},\ }\href {\doibase https://doi.org/10.1016/0304-8853(87)90721-9}
  {\bibfield  {journal} {\bibinfo  {journal} {J. Magn. Magn. Mater.}\ }\textbf
  {\bibinfo {volume} {67}},\ \bibinfo {pages} {65} (\bibinfo {year}
  {1987})}\BibitemShut {NoStop}%
\bibitem [{\citenamefont {Pajda}\ \emph {et~al.}(2001)\citenamefont {Pajda},
  \citenamefont {Kudrnovsk\'y}, \citenamefont {Turek}, \citenamefont {Drchal},\
  and\ \citenamefont {Bruno}}]{Pajda01}%
  \BibitemOpen
  \bibfield  {author} {\bibinfo {author} {\bibfnamefont {M.}~\bibnamefont
  {Pajda}}, \bibinfo {author} {\bibfnamefont {J.}~\bibnamefont {Kudrnovsk\'y}},
  \bibinfo {author} {\bibfnamefont {I.}~\bibnamefont {Turek}}, \bibinfo
  {author} {\bibfnamefont {V.}~\bibnamefont {Drchal}}, \ and\ \bibinfo {author}
  {\bibfnamefont {P.}~\bibnamefont {Bruno}},\ }\href {\doibase
  10.1103/PhysRevB.64.174402} {\bibfield  {journal} {\bibinfo  {journal} {Phys.
  Rev. B}\ }\textbf {\bibinfo {volume} {64}},\ \bibinfo {pages} {174402}
  (\bibinfo {year} {2001})}\BibitemShut {NoStop}%
\bibitem [{\citenamefont {LaShell}\ \emph {et~al.}(1996)\citenamefont
  {LaShell}, \citenamefont {McDougall},\ and\ \citenamefont
  {Jensen}}]{LaShell96}%
  \BibitemOpen
  \bibfield  {author} {\bibinfo {author} {\bibfnamefont {S.}~\bibnamefont
  {LaShell}}, \bibinfo {author} {\bibfnamefont {B.~A.}\ \bibnamefont
  {McDougall}}, \ and\ \bibinfo {author} {\bibfnamefont {E.}~\bibnamefont
  {Jensen}},\ }\href {\doibase 10.1103/PhysRevLett.77.3419} {\bibfield
  {journal} {\bibinfo  {journal} {Phys. Rev. Lett.}\ }\textbf {\bibinfo
  {volume} {77}},\ \bibinfo {pages} {3419} (\bibinfo {year}
  {1996})}\BibitemShut {NoStop}%
\bibitem [{\citenamefont {Hoesch}\ \emph {et~al.}(2004)\citenamefont {Hoesch},
  \citenamefont {Muntwiler}, \citenamefont {Petrov}, \citenamefont
  {Hengsberger}, \citenamefont {Patthey}, \citenamefont {Shi}, \citenamefont
  {Falub}, \citenamefont {Greber},\ and\ \citenamefont
  {Osterwalder}}]{Hoesch04}%
  \BibitemOpen
  \bibfield  {author} {\bibinfo {author} {\bibfnamefont {M.}~\bibnamefont
  {Hoesch}}, \bibinfo {author} {\bibfnamefont {M.}~\bibnamefont {Muntwiler}},
  \bibinfo {author} {\bibfnamefont {V.~N.}\ \bibnamefont {Petrov}}, \bibinfo
  {author} {\bibfnamefont {M.}~\bibnamefont {Hengsberger}}, \bibinfo {author}
  {\bibfnamefont {L.}~\bibnamefont {Patthey}}, \bibinfo {author} {\bibfnamefont
  {M.}~\bibnamefont {Shi}}, \bibinfo {author} {\bibfnamefont {M.}~\bibnamefont
  {Falub}}, \bibinfo {author} {\bibfnamefont {T.}~\bibnamefont {Greber}}, \
  and\ \bibinfo {author} {\bibfnamefont {J.}~\bibnamefont {Osterwalder}},\
  }\href {\doibase 10.1103/PhysRevB.69.241401} {\bibfield  {journal} {\bibinfo
  {journal} {Phys. Rev. B}\ }\textbf {\bibinfo {volume} {69}},\ \bibinfo
  {pages} {241401} (\bibinfo {year} {2004})}\BibitemShut {NoStop}%
\bibitem [{\citenamefont {Koroteev}\ \emph {et~al.}(2004)\citenamefont
  {Koroteev}, \citenamefont {Bihlmayer}, \citenamefont {Gayone}, \citenamefont
  {Chulkov}, \citenamefont {Bl\"ugel}, \citenamefont {Echenique},\ and\
  \citenamefont {Hofmann}}]{Koroteev04}%
  \BibitemOpen
  \bibfield  {author} {\bibinfo {author} {\bibfnamefont {Y.~M.}\ \bibnamefont
  {Koroteev}}, \bibinfo {author} {\bibfnamefont {G.}~\bibnamefont {Bihlmayer}},
  \bibinfo {author} {\bibfnamefont {J.~E.}\ \bibnamefont {Gayone}}, \bibinfo
  {author} {\bibfnamefont {E.~V.}\ \bibnamefont {Chulkov}}, \bibinfo {author}
  {\bibfnamefont {S.}~\bibnamefont {Bl\"ugel}}, \bibinfo {author}
  {\bibfnamefont {P.~M.}\ \bibnamefont {Echenique}}, \ and\ \bibinfo {author}
  {\bibfnamefont {P.}~\bibnamefont {Hofmann}},\ }\href {\doibase
  10.1103/PhysRevLett.93.046403} {\bibfield  {journal} {\bibinfo  {journal}
  {Phys. Rev. Lett.}\ }\textbf {\bibinfo {volume} {93}},\ \bibinfo {pages}
  {046403} (\bibinfo {year} {2004})}\BibitemShut {NoStop}%
\bibitem [{\citenamefont {Ast}\ \emph {et~al.}(2007)\citenamefont {Ast},
  \citenamefont {Henk}, \citenamefont {Ernst}, \citenamefont {Moreschini},
  \citenamefont {Falub}, \citenamefont {Pacil\'e}, \citenamefont {Bruno},
  \citenamefont {Kern},\ and\ \citenamefont {Grioni}}]{Ast07}%
  \BibitemOpen
  \bibfield  {author} {\bibinfo {author} {\bibfnamefont {C.~R.}\ \bibnamefont
  {Ast}}, \bibinfo {author} {\bibfnamefont {J.}~\bibnamefont {Henk}}, \bibinfo
  {author} {\bibfnamefont {A.}~\bibnamefont {Ernst}}, \bibinfo {author}
  {\bibfnamefont {L.}~\bibnamefont {Moreschini}}, \bibinfo {author}
  {\bibfnamefont {M.~C.}\ \bibnamefont {Falub}}, \bibinfo {author}
  {\bibfnamefont {D.}~\bibnamefont {Pacil\'e}}, \bibinfo {author}
  {\bibfnamefont {P.}~\bibnamefont {Bruno}}, \bibinfo {author} {\bibfnamefont
  {K.}~\bibnamefont {Kern}}, \ and\ \bibinfo {author} {\bibfnamefont
  {M.}~\bibnamefont {Grioni}},\ }\href {\doibase 10.1103/PhysRevLett.98.186807}
  {\bibfield  {journal} {\bibinfo  {journal} {Phys. Rev. Lett.}\ }\textbf
  {\bibinfo {volume} {98}},\ \bibinfo {pages} {186807} (\bibinfo {year}
  {2007})}\BibitemShut {NoStop}%
\bibitem [{\citenamefont {He}\ \emph {et~al.}(2008)\citenamefont {He},
  \citenamefont {Hirahara}, \citenamefont {Okuda}, \citenamefont {Hasegawa},
  \citenamefont {Kakizaki},\ and\ \citenamefont {Matsuda}}]{He08}%
  \BibitemOpen
  \bibfield  {author} {\bibinfo {author} {\bibfnamefont {K.}~\bibnamefont
  {He}}, \bibinfo {author} {\bibfnamefont {T.}~\bibnamefont {Hirahara}},
  \bibinfo {author} {\bibfnamefont {T.}~\bibnamefont {Okuda}}, \bibinfo
  {author} {\bibfnamefont {S.}~\bibnamefont {Hasegawa}}, \bibinfo {author}
  {\bibfnamefont {A.}~\bibnamefont {Kakizaki}}, \ and\ \bibinfo {author}
  {\bibfnamefont {I.}~\bibnamefont {Matsuda}},\ }\href {\doibase
  10.1103/PhysRevLett.101.107604} {\bibfield  {journal} {\bibinfo  {journal}
  {Phys. Rev. Lett.}\ }\textbf {\bibinfo {volume} {101}},\ \bibinfo {pages}
  {107604} (\bibinfo {year} {2008})}\BibitemShut {NoStop}%
\bibitem [{\citenamefont {Frantzeskakis}\ \emph {et~al.}(2008)\citenamefont
  {Frantzeskakis}, \citenamefont {Pons}, \citenamefont {Mirhosseini},
  \citenamefont {Henk}, \citenamefont {Ast},\ and\ \citenamefont
  {Grioni}}]{Frantzeskakis08}%
  \BibitemOpen
  \bibfield  {author} {\bibinfo {author} {\bibfnamefont {E.}~\bibnamefont
  {Frantzeskakis}}, \bibinfo {author} {\bibfnamefont {S.}~\bibnamefont {Pons}},
  \bibinfo {author} {\bibfnamefont {H.}~\bibnamefont {Mirhosseini}}, \bibinfo
  {author} {\bibfnamefont {J.}~\bibnamefont {Henk}}, \bibinfo {author}
  {\bibfnamefont {C.~R.}\ \bibnamefont {Ast}}, \ and\ \bibinfo {author}
  {\bibfnamefont {M.}~\bibnamefont {Grioni}},\ }\href {\doibase
  10.1103/PhysRevLett.101.196805} {\bibfield  {journal} {\bibinfo  {journal}
  {Phys. Rev. Lett.}\ }\textbf {\bibinfo {volume} {101}},\ \bibinfo {pages}
  {196805} (\bibinfo {year} {2008})}\BibitemShut {NoStop}%
\bibitem [{\citenamefont {King}\ \emph {et~al.}(2011)\citenamefont {King},
  \citenamefont {Hatch}, \citenamefont {Bianchi}, \citenamefont {Ovsyannikov},
  \citenamefont {Lupulescu}, \citenamefont {Landolt}, \citenamefont {Slomski},
  \citenamefont {Dil}, \citenamefont {Guan}, \citenamefont {Mi}, \citenamefont
  {Rienks}, \citenamefont {Fink}, \citenamefont {Lindblad}, \citenamefont
  {Svensson}, \citenamefont {Bao}, \citenamefont {Balakrishnan}, \citenamefont
  {Iversen}, \citenamefont {Osterwalder}, \citenamefont {Eberhardt},
  \citenamefont {Baumberger},\ and\ \citenamefont {Hofmann}}]{King11}%
  \BibitemOpen
  \bibfield  {author} {\bibinfo {author} {\bibfnamefont {P.~D.~C.}\
  \bibnamefont {King}}, \bibinfo {author} {\bibfnamefont {R.~C.}\ \bibnamefont
  {Hatch}}, \bibinfo {author} {\bibfnamefont {M.}~\bibnamefont {Bianchi}},
  \bibinfo {author} {\bibfnamefont {R.}~\bibnamefont {Ovsyannikov}}, \bibinfo
  {author} {\bibfnamefont {C.}~\bibnamefont {Lupulescu}}, \bibinfo {author}
  {\bibfnamefont {G.}~\bibnamefont {Landolt}}, \bibinfo {author} {\bibfnamefont
  {B.}~\bibnamefont {Slomski}}, \bibinfo {author} {\bibfnamefont {J.~H.}\
  \bibnamefont {Dil}}, \bibinfo {author} {\bibfnamefont {D.}~\bibnamefont
  {Guan}}, \bibinfo {author} {\bibfnamefont {J.~L.}\ \bibnamefont {Mi}},
  \bibinfo {author} {\bibfnamefont {E.~D.~L.}\ \bibnamefont {Rienks}}, \bibinfo
  {author} {\bibfnamefont {J.}~\bibnamefont {Fink}}, \bibinfo {author}
  {\bibfnamefont {A.}~\bibnamefont {Lindblad}}, \bibinfo {author}
  {\bibfnamefont {S.}~\bibnamefont {Svensson}}, \bibinfo {author}
  {\bibfnamefont {S.}~\bibnamefont {Bao}}, \bibinfo {author} {\bibfnamefont
  {G.}~\bibnamefont {Balakrishnan}}, \bibinfo {author} {\bibfnamefont {B.~B.}\
  \bibnamefont {Iversen}}, \bibinfo {author} {\bibfnamefont {J.}~\bibnamefont
  {Osterwalder}}, \bibinfo {author} {\bibfnamefont {W.}~\bibnamefont
  {Eberhardt}}, \bibinfo {author} {\bibfnamefont {F.}~\bibnamefont
  {Baumberger}}, \ and\ \bibinfo {author} {\bibfnamefont {P.}~\bibnamefont
  {Hofmann}},\ }\href {\doibase 10.1103/PhysRevLett.107.096802} {\bibfield
  {journal} {\bibinfo  {journal} {Phys. Rev. Lett.}\ }\textbf {\bibinfo
  {volume} {107}},\ \bibinfo {pages} {096802} (\bibinfo {year}
  {2011})}\BibitemShut {NoStop}%
\bibitem [{\citenamefont {Zhu}\ \emph {et~al.}(2011)\citenamefont {Zhu},
  \citenamefont {Levy}, \citenamefont {Ludbrook}, \citenamefont {Veenstra},
  \citenamefont {Rosen}, \citenamefont {Comin}, \citenamefont {Wong},
  \citenamefont {Dosanjh}, \citenamefont {Ubaldini}, \citenamefont {Syers},
  \citenamefont {Butch}, \citenamefont {Paglione}, \citenamefont {Elfimov},\
  and\ \citenamefont {Damascelli}}]{Zhu11}%
  \BibitemOpen
  \bibfield  {author} {\bibinfo {author} {\bibfnamefont {Z.-H.}\ \bibnamefont
  {Zhu}}, \bibinfo {author} {\bibfnamefont {G.}~\bibnamefont {Levy}}, \bibinfo
  {author} {\bibfnamefont {B.}~\bibnamefont {Ludbrook}}, \bibinfo {author}
  {\bibfnamefont {C.~N.}\ \bibnamefont {Veenstra}}, \bibinfo {author}
  {\bibfnamefont {J.~A.}\ \bibnamefont {Rosen}}, \bibinfo {author}
  {\bibfnamefont {R.}~\bibnamefont {Comin}}, \bibinfo {author} {\bibfnamefont
  {D.}~\bibnamefont {Wong}}, \bibinfo {author} {\bibfnamefont {P.}~\bibnamefont
  {Dosanjh}}, \bibinfo {author} {\bibfnamefont {A.}~\bibnamefont {Ubaldini}},
  \bibinfo {author} {\bibfnamefont {P.}~\bibnamefont {Syers}}, \bibinfo
  {author} {\bibfnamefont {N.~P.}\ \bibnamefont {Butch}}, \bibinfo {author}
  {\bibfnamefont {J.}~\bibnamefont {Paglione}}, \bibinfo {author}
  {\bibfnamefont {I.~S.}\ \bibnamefont {Elfimov}}, \ and\ \bibinfo {author}
  {\bibfnamefont {A.}~\bibnamefont {Damascelli}},\ }\href {\doibase
  10.1103/PhysRevLett.107.186405} {\bibfield  {journal} {\bibinfo  {journal}
  {Phys. Rev. Lett.}\ }\textbf {\bibinfo {volume} {107}},\ \bibinfo {pages}
  {186405} (\bibinfo {year} {2011})}\BibitemShut {NoStop}%
\bibitem [{\citenamefont {Zegarra}\ \emph {et~al.}(2020)\citenamefont
  {Zegarra}, \citenamefont {Egues},\ and\ \citenamefont {Chen}}]{Zegarra20}%
  \BibitemOpen
  \bibfield  {author} {\bibinfo {author} {\bibfnamefont {A.}~\bibnamefont
  {Zegarra}}, \bibinfo {author} {\bibfnamefont {J.~C.}\ \bibnamefont {Egues}},
  \ and\ \bibinfo {author} {\bibfnamefont {W.}~\bibnamefont {Chen}},\ }\href
  {\doibase 10.1103/PhysRevB.101.224438} {\bibfield  {journal} {\bibinfo
  {journal} {Phys. Rev. B}\ }\textbf {\bibinfo {volume} {101}},\ \bibinfo
  {pages} {224438} (\bibinfo {year} {2020})}\BibitemShut {NoStop}%
\bibitem [{\citenamefont {Chen}(2020{\natexlab{a}})}]{Chen20_TIFMM}%
  \BibitemOpen
  \bibfield  {author} {\bibinfo {author} {\bibfnamefont {W.}~\bibnamefont
  {Chen}},\ }\href {\doibase 10.1103/PhysRevB.102.144442} {\bibfield  {journal}
  {\bibinfo  {journal} {Phys. Rev. B}\ }\textbf {\bibinfo {volume} {102}},\
  \bibinfo {pages} {144442} (\bibinfo {year} {2020}{\natexlab{a}})}\BibitemShut
  {NoStop}%
\bibitem [{\citenamefont {Rashba}(2003)}]{Rashba03}%
  \BibitemOpen
  \bibfield  {author} {\bibinfo {author} {\bibfnamefont {E.~I.}\ \bibnamefont
  {Rashba}},\ }\href {\doibase 10.1103/PhysRevB.68.241315} {\bibfield
  {journal} {\bibinfo  {journal} {Phys. Rev. B}\ }\textbf {\bibinfo {volume}
  {68}},\ \bibinfo {pages} {241315} (\bibinfo {year} {2003})}\BibitemShut
  {NoStop}%
\bibitem [{\citenamefont {Rashba}(2005)}]{Rashba05}%
  \BibitemOpen
  \bibfield  {author} {\bibinfo {author} {\bibfnamefont {E.~I.}\ \bibnamefont
  {Rashba}},\ }\href {\doibase 10.1007/s10948-005-3349-8} {\bibfield  {journal}
  {\bibinfo  {journal} {J. Supercond.}\ }\textbf {\bibinfo {volume} {18}},\
  \bibinfo {pages} {137} (\bibinfo {year} {2005})}\BibitemShut {NoStop}%
\bibitem [{\citenamefont {Sonin}(2007)}]{Sonin07}%
  \BibitemOpen
  \bibfield  {author} {\bibinfo {author} {\bibfnamefont {E.~B.}\ \bibnamefont
  {Sonin}},\ }\href {\doibase 10.1103/PhysRevB.76.033306} {\bibfield  {journal}
  {\bibinfo  {journal} {Phys. Rev. B}\ }\textbf {\bibinfo {volume} {76}},\
  \bibinfo {pages} {033306} (\bibinfo {year} {2007})}\BibitemShut {NoStop}%
\bibitem [{\citenamefont {Tokatly}(2008)}]{Tokatly08}%
  \BibitemOpen
  \bibfield  {author} {\bibinfo {author} {\bibfnamefont {I.~V.}\ \bibnamefont
  {Tokatly}},\ }\href {\doibase 10.1103/PhysRevLett.101.106601} {\bibfield
  {journal} {\bibinfo  {journal} {Phys. Rev. Lett.}\ }\textbf {\bibinfo
  {volume} {101}},\ \bibinfo {pages} {106601} (\bibinfo {year}
  {2008})}\BibitemShut {NoStop}%
\bibitem [{\citenamefont {Kane}\ and\ \citenamefont {Mele}(2005)}]{Kane05_2}%
  \BibitemOpen
  \bibfield  {author} {\bibinfo {author} {\bibfnamefont {C.~L.}\ \bibnamefont
  {Kane}}\ and\ \bibinfo {author} {\bibfnamefont {E.~J.}\ \bibnamefont
  {Mele}},\ }\href {\doibase 10.1103/PhysRevLett.95.226801} {\bibfield
  {journal} {\bibinfo  {journal} {Phys. Rev. Lett.}\ }\textbf {\bibinfo
  {volume} {95}},\ \bibinfo {pages} {226801} (\bibinfo {year}
  {2005})}\BibitemShut {NoStop}%
\bibitem [{\citenamefont {Bernevig}\ \emph {et~al.}(2006)\citenamefont
  {Bernevig}, \citenamefont {Hughes},\ and\ \citenamefont
  {Zhang}}]{Bernevig06}%
  \BibitemOpen
  \bibfield  {author} {\bibinfo {author} {\bibfnamefont {B.~A.}\ \bibnamefont
  {Bernevig}}, \bibinfo {author} {\bibfnamefont {T.~L.}\ \bibnamefont
  {Hughes}}, \ and\ \bibinfo {author} {\bibfnamefont {S.-C.}\ \bibnamefont
  {Zhang}},\ }\href {\doibase 10.1126/science.1133734} {\bibfield  {journal}
  {\bibinfo  {journal} {Science}\ }\textbf {\bibinfo {volume} {314}},\ \bibinfo
  {pages} {1757} (\bibinfo {year} {2006})}\BibitemShut {NoStop}%
\bibitem [{\citenamefont {Bernevig}\ and\ \citenamefont
  {Zhang}(2006)}]{Bernevig06_2}%
  \BibitemOpen
  \bibfield  {author} {\bibinfo {author} {\bibfnamefont {B.~A.}\ \bibnamefont
  {Bernevig}}\ and\ \bibinfo {author} {\bibfnamefont {S.-C.}\ \bibnamefont
  {Zhang}},\ }\href {\doibase 10.1103/PhysRevLett.96.106802} {\bibfield
  {journal} {\bibinfo  {journal} {Phys. Rev. Lett.}\ }\textbf {\bibinfo
  {volume} {96}},\ \bibinfo {pages} {106802} (\bibinfo {year}
  {2006})}\BibitemShut {NoStop}%
\bibitem [{\citenamefont {K{\"o}nig}\ \emph {et~al.}(2007)\citenamefont
  {K{\"o}nig}, \citenamefont {Wiedmann}, \citenamefont {Br{\"u}ne},
  \citenamefont {Roth}, \citenamefont {Buhmann}, \citenamefont {Molenkamp},
  \citenamefont {Qi},\ and\ \citenamefont {Zhang}}]{Konig07}%
  \BibitemOpen
  \bibfield  {author} {\bibinfo {author} {\bibfnamefont {M.}~\bibnamefont
  {K{\"o}nig}}, \bibinfo {author} {\bibfnamefont {S.}~\bibnamefont {Wiedmann}},
  \bibinfo {author} {\bibfnamefont {C.}~\bibnamefont {Br{\"u}ne}}, \bibinfo
  {author} {\bibfnamefont {A.}~\bibnamefont {Roth}}, \bibinfo {author}
  {\bibfnamefont {H.}~\bibnamefont {Buhmann}}, \bibinfo {author} {\bibfnamefont
  {L.~W.}\ \bibnamefont {Molenkamp}}, \bibinfo {author} {\bibfnamefont {X.-L.}\
  \bibnamefont {Qi}}, \ and\ \bibinfo {author} {\bibfnamefont {S.-C.}\
  \bibnamefont {Zhang}},\ }\href {\doibase 10.1126/science.1148047} {\bibfield
  {journal} {\bibinfo  {journal} {Science}\ }\textbf {\bibinfo {volume}
  {318}},\ \bibinfo {pages} {766} (\bibinfo {year} {2007})}\BibitemShut
  {NoStop}%
\bibitem [{\citenamefont
  {Chen}(2020{\natexlab{b}})}]{Chen20_absence_edge_current}%
  \BibitemOpen
  \bibfield  {author} {\bibinfo {author} {\bibfnamefont {W.}~\bibnamefont
  {Chen}},\ }\href {\doibase 10.1103/PhysRevB.101.195120} {\bibfield  {journal}
  {\bibinfo  {journal} {Phys. Rev. B}\ }\textbf {\bibinfo {volume} {101}},\
  \bibinfo {pages} {195120} (\bibinfo {year} {2020}{\natexlab{b}})}\BibitemShut
  {NoStop}%
\bibitem [{\citenamefont {Schultz}\ \emph {et~al.}(1996)\citenamefont
  {Schultz}, \citenamefont {Heinrichs}, \citenamefont {Merkt}, \citenamefont
  {Colin}, \citenamefont {Skauli},\ and\ \citenamefont
  {L{\o}vold}}]{Schultz96}%
  \BibitemOpen
  \bibfield  {author} {\bibinfo {author} {\bibfnamefont {M.}~\bibnamefont
  {Schultz}}, \bibinfo {author} {\bibfnamefont {F.}~\bibnamefont {Heinrichs}},
  \bibinfo {author} {\bibfnamefont {U.}~\bibnamefont {Merkt}}, \bibinfo
  {author} {\bibfnamefont {T.}~\bibnamefont {Colin}}, \bibinfo {author}
  {\bibfnamefont {T.}~\bibnamefont {Skauli}}, \ and\ \bibinfo {author}
  {\bibfnamefont {S.}~\bibnamefont {L{\o}vold}},\ }\href {\doibase
  10.1088/0268-1242/11/8/009} {\bibfield  {journal} {\bibinfo  {journal}
  {Semicond. Sci. Technol.}\ }\textbf {\bibinfo {volume} {11}},\ \bibinfo
  {pages} {1168} (\bibinfo {year} {1996})}\BibitemShut {NoStop}%
\bibitem [{\citenamefont {Nitta}\ \emph {et~al.}(1997)\citenamefont {Nitta},
  \citenamefont {Akazaki}, \citenamefont {Takayanagi},\ and\ \citenamefont
  {Enoki}}]{Nitta97}%
  \BibitemOpen
  \bibfield  {author} {\bibinfo {author} {\bibfnamefont {J.}~\bibnamefont
  {Nitta}}, \bibinfo {author} {\bibfnamefont {T.}~\bibnamefont {Akazaki}},
  \bibinfo {author} {\bibfnamefont {H.}~\bibnamefont {Takayanagi}}, \ and\
  \bibinfo {author} {\bibfnamefont {T.}~\bibnamefont {Enoki}},\ }\href
  {\doibase 10.1103/PhysRevLett.78.1335} {\bibfield  {journal} {\bibinfo
  {journal} {Phys. Rev. Lett.}\ }\textbf {\bibinfo {volume} {78}},\ \bibinfo
  {pages} {1335} (\bibinfo {year} {1997})}\BibitemShut {NoStop}%
\bibitem [{\citenamefont {Nitta}\ \emph {et~al.}(1998)\citenamefont {Nitta},
  \citenamefont {Akazaki}, \citenamefont {Takayanagi},\ and\ \citenamefont
  {Enoki}}]{Nitta98}%
  \BibitemOpen
  \bibfield  {author} {\bibinfo {author} {\bibfnamefont {J.}~\bibnamefont
  {Nitta}}, \bibinfo {author} {\bibfnamefont {T.}~\bibnamefont {Akazaki}},
  \bibinfo {author} {\bibfnamefont {H.}~\bibnamefont {Takayanagi}}, \ and\
  \bibinfo {author} {\bibfnamefont {T.}~\bibnamefont {Enoki}},\ }\href
  {\doibase https://doi.org/10.1016/S1386-9477(98)00109-X} {\bibfield
  {journal} {\bibinfo  {journal} {Physica E}\ }\textbf {\bibinfo {volume}
  {2}},\ \bibinfo {pages} {527} (\bibinfo {year} {1998})}\BibitemShut {NoStop}%
\bibitem [{\citenamefont {Heida}\ \emph {et~al.}(1998)\citenamefont {Heida},
  \citenamefont {van Wees}, \citenamefont {Kuipers}, \citenamefont {Klapwijk},\
  and\ \citenamefont {Borghs}}]{Heida98}%
  \BibitemOpen
  \bibfield  {author} {\bibinfo {author} {\bibfnamefont {J.~P.}\ \bibnamefont
  {Heida}}, \bibinfo {author} {\bibfnamefont {B.~J.}\ \bibnamefont {van Wees}},
  \bibinfo {author} {\bibfnamefont {J.~J.}\ \bibnamefont {Kuipers}}, \bibinfo
  {author} {\bibfnamefont {T.~M.}\ \bibnamefont {Klapwijk}}, \ and\ \bibinfo
  {author} {\bibfnamefont {G.}~\bibnamefont {Borghs}},\ }\href {\doibase
  10.1103/PhysRevB.57.11911} {\bibfield  {journal} {\bibinfo  {journal} {Phys.
  Rev. B}\ }\textbf {\bibinfo {volume} {57}},\ \bibinfo {pages} {11911}
  (\bibinfo {year} {1998})}\BibitemShut {NoStop}%
\bibitem [{\citenamefont {Miller}\ \emph {et~al.}(2003)\citenamefont {Miller},
  \citenamefont {Zumb\"uhl}, \citenamefont {Marcus}, \citenamefont
  {Lyanda-Geller}, \citenamefont {Goldhaber-Gordon}, \citenamefont {Campman},\
  and\ \citenamefont {Gossard}}]{Miller03}%
  \BibitemOpen
  \bibfield  {author} {\bibinfo {author} {\bibfnamefont {J.~B.}\ \bibnamefont
  {Miller}}, \bibinfo {author} {\bibfnamefont {D.~M.}\ \bibnamefont
  {Zumb\"uhl}}, \bibinfo {author} {\bibfnamefont {C.~M.}\ \bibnamefont
  {Marcus}}, \bibinfo {author} {\bibfnamefont {Y.~B.}\ \bibnamefont
  {Lyanda-Geller}}, \bibinfo {author} {\bibfnamefont {D.}~\bibnamefont
  {Goldhaber-Gordon}}, \bibinfo {author} {\bibfnamefont {K.}~\bibnamefont
  {Campman}}, \ and\ \bibinfo {author} {\bibfnamefont {A.~C.}\ \bibnamefont
  {Gossard}},\ }\href {\doibase 10.1103/PhysRevLett.90.076807} {\bibfield
  {journal} {\bibinfo  {journal} {Phys. Rev. Lett.}\ }\textbf {\bibinfo
  {volume} {90}},\ \bibinfo {pages} {076807} (\bibinfo {year}
  {2003})}\BibitemShut {NoStop}%
\bibitem [{\citenamefont {Hinz}\ \emph {et~al.}(2006)\citenamefont {Hinz},
  \citenamefont {Buhmann}, \citenamefont {Schäfer}, \citenamefont {Hock},
  \citenamefont {Becker},\ and\ \citenamefont {Molenkamp}}]{Hinz06}%
  \BibitemOpen
  \bibfield  {author} {\bibinfo {author} {\bibfnamefont {J.}~\bibnamefont
  {Hinz}}, \bibinfo {author} {\bibfnamefont {H.}~\bibnamefont {Buhmann}},
  \bibinfo {author} {\bibfnamefont {M.}~\bibnamefont {Schäfer}}, \bibinfo
  {author} {\bibfnamefont {V.}~\bibnamefont {Hock}}, \bibinfo {author}
  {\bibfnamefont {C.~R.}\ \bibnamefont {Becker}}, \ and\ \bibinfo {author}
  {\bibfnamefont {L.~W.}\ \bibnamefont {Molenkamp}},\ }\href {\doibase
  10.1088/0268-1242/21/4/015} {\bibfield  {journal} {\bibinfo  {journal}
  {Semicond. Sci. Technol.}\ }\textbf {\bibinfo {volume} {21}},\ \bibinfo
  {pages} {501} (\bibinfo {year} {2006})}\BibitemShut {NoStop}%
\bibitem [{\citenamefont {Studer}\ \emph {et~al.}(2009)\citenamefont {Studer},
  \citenamefont {Salis}, \citenamefont {Ensslin}, \citenamefont {Driscoll},\
  and\ \citenamefont {Gossard}}]{Studer09}%
  \BibitemOpen
  \bibfield  {author} {\bibinfo {author} {\bibfnamefont {M.}~\bibnamefont
  {Studer}}, \bibinfo {author} {\bibfnamefont {G.}~\bibnamefont {Salis}},
  \bibinfo {author} {\bibfnamefont {K.}~\bibnamefont {Ensslin}}, \bibinfo
  {author} {\bibfnamefont {D.~C.}\ \bibnamefont {Driscoll}}, \ and\ \bibinfo
  {author} {\bibfnamefont {A.~C.}\ \bibnamefont {Gossard}},\ }\href {\doibase
  10.1103/PhysRevLett.103.027201} {\bibfield  {journal} {\bibinfo  {journal}
  {Phys. Rev. Lett.}\ }\textbf {\bibinfo {volume} {103}},\ \bibinfo {pages}
  {027201} (\bibinfo {year} {2009})}\BibitemShut {NoStop}%
\bibitem [{\citenamefont {Koo}\ \emph {et~al.}(2009)\citenamefont {Koo},
  \citenamefont {Kwon}, \citenamefont {Eom}, \citenamefont {Chang},
  \citenamefont {Han},\ and\ \citenamefont {Johnson}}]{Koo09}%
  \BibitemOpen
  \bibfield  {author} {\bibinfo {author} {\bibfnamefont {H.~C.}\ \bibnamefont
  {Koo}}, \bibinfo {author} {\bibfnamefont {J.~H.}\ \bibnamefont {Kwon}},
  \bibinfo {author} {\bibfnamefont {J.}~\bibnamefont {Eom}}, \bibinfo {author}
  {\bibfnamefont {J.}~\bibnamefont {Chang}}, \bibinfo {author} {\bibfnamefont
  {S.~H.}\ \bibnamefont {Han}}, \ and\ \bibinfo {author} {\bibfnamefont
  {M.}~\bibnamefont {Johnson}},\ }\href {\doibase 10.1126/science.1173667}
  {\bibfield  {journal} {\bibinfo  {journal} {Science}\ }\textbf {\bibinfo
  {volume} {325}},\ \bibinfo {pages} {1515} (\bibinfo {year}
  {2009})}\BibitemShut {NoStop}%
\bibitem [{\citenamefont {Liang}\ and\ \citenamefont {Gao}(2012)}]{Liang12}%
  \BibitemOpen
  \bibfield  {author} {\bibinfo {author} {\bibfnamefont {D.}~\bibnamefont
  {Liang}}\ and\ \bibinfo {author} {\bibfnamefont {X.~P.}\ \bibnamefont
  {Gao}},\ }\href {\doibase 10.1021/nl301325h} {\bibfield  {journal} {\bibinfo
  {journal} {Nano Lett.}\ }\textbf {\bibinfo {volume} {12}},\ \bibinfo {pages}
  {3263} (\bibinfo {year} {2012})}\BibitemShut {NoStop}%
\bibitem [{\citenamefont {Takase}\ \emph {et~al.}(2017)\citenamefont {Takase},
  \citenamefont {Ashikawa}, \citenamefont {Zhang}, \citenamefont {Tateno},\
  and\ \citenamefont {Sasaki}}]{Takase17}%
  \BibitemOpen
  \bibfield  {author} {\bibinfo {author} {\bibfnamefont {K.}~\bibnamefont
  {Takase}}, \bibinfo {author} {\bibfnamefont {Y.}~\bibnamefont {Ashikawa}},
  \bibinfo {author} {\bibfnamefont {G.}~\bibnamefont {Zhang}}, \bibinfo
  {author} {\bibfnamefont {K.}~\bibnamefont {Tateno}}, \ and\ \bibinfo {author}
  {\bibfnamefont {S.}~\bibnamefont {Sasaki}},\ }\href {\doibase
  10.1038/s41598-017-01080-0} {\bibfield  {journal} {\bibinfo  {journal} {Sci.
  Rep.}\ }\textbf {\bibinfo {volume} {7}},\ \bibinfo {pages} {930} (\bibinfo
  {year} {2017})}\BibitemShut {NoStop}%
\bibitem [{\citenamefont {Chen}\ \emph
  {et~al.}(2018{\natexlab{a}})\citenamefont {Chen}, \citenamefont {Gmitra},
  \citenamefont {Vogel}, \citenamefont {Islinger}, \citenamefont {Kronseder},
  \citenamefont {Schuh}, \citenamefont {Bougeard}, \citenamefont {Fabian},
  \citenamefont {Weiss},\ and\ \citenamefont {Back}}]{Chen18_tunable_SOC}%
  \BibitemOpen
  \bibfield  {author} {\bibinfo {author} {\bibfnamefont {L.}~\bibnamefont
  {Chen}}, \bibinfo {author} {\bibfnamefont {M.}~\bibnamefont {Gmitra}},
  \bibinfo {author} {\bibfnamefont {M.}~\bibnamefont {Vogel}}, \bibinfo
  {author} {\bibfnamefont {R.}~\bibnamefont {Islinger}}, \bibinfo {author}
  {\bibfnamefont {M.}~\bibnamefont {Kronseder}}, \bibinfo {author}
  {\bibfnamefont {D.}~\bibnamefont {Schuh}}, \bibinfo {author} {\bibfnamefont
  {D.}~\bibnamefont {Bougeard}}, \bibinfo {author} {\bibfnamefont
  {J.}~\bibnamefont {Fabian}}, \bibinfo {author} {\bibfnamefont
  {D.}~\bibnamefont {Weiss}}, \ and\ \bibinfo {author} {\bibfnamefont {C.~H.}\
  \bibnamefont {Back}},\ }\href {\doibase 10.1038/s41928-018-0085-1} {\bibfield
   {journal} {\bibinfo  {journal} {Nat. Electron.}\ }\textbf {\bibinfo {volume}
  {1}},\ \bibinfo {pages} {350} (\bibinfo {year}
  {2018}{\natexlab{a}})}\BibitemShut {NoStop}%
\bibitem [{\citenamefont {Froning}\ \emph {et~al.}(2021)\citenamefont
  {Froning}, \citenamefont {Camenzind}, \citenamefont {van~der Molen},
  \citenamefont {Li}, \citenamefont {Bakkers}, \citenamefont {Zumb\"{u}hl},\
  and\ \citenamefont {Braakman}}]{Froning21}%
  \BibitemOpen
  \bibfield  {author} {\bibinfo {author} {\bibfnamefont {F.~N.~M.}\
  \bibnamefont {Froning}}, \bibinfo {author} {\bibfnamefont {L.~C.}\
  \bibnamefont {Camenzind}}, \bibinfo {author} {\bibfnamefont {O.~A.~H.}\
  \bibnamefont {van~der Molen}}, \bibinfo {author} {\bibfnamefont
  {A.}~\bibnamefont {Li}}, \bibinfo {author} {\bibfnamefont {E.~P. A.~M.}\
  \bibnamefont {Bakkers}}, \bibinfo {author} {\bibfnamefont {D.~M.}\
  \bibnamefont {Zumb\"{u}hl}}, \ and\ \bibinfo {author} {\bibfnamefont {F.~R.}\
  \bibnamefont {Braakman}},\ }\href {\doibase 10.1038/s41565-020-00828-6}
  {\bibfield  {journal} {\bibinfo  {journal} {Nat. Nanotechnol.}\ }\textbf
  {\bibinfo {volume} {16}},\ \bibinfo {pages} {308} (\bibinfo {year}
  {2021})}\BibitemShut {NoStop}%
\bibitem [{\citenamefont {Dushenko}\ \emph {et~al.}(2018)\citenamefont
  {Dushenko}, \citenamefont {Hokazono}, \citenamefont {Nakamura}, \citenamefont
  {Ando}, \citenamefont {Shinjo},\ and\ \citenamefont
  {Shiraishi}}]{Dushenko18}%
  \BibitemOpen
  \bibfield  {author} {\bibinfo {author} {\bibfnamefont {S.}~\bibnamefont
  {Dushenko}}, \bibinfo {author} {\bibfnamefont {M.}~\bibnamefont {Hokazono}},
  \bibinfo {author} {\bibfnamefont {K.}~\bibnamefont {Nakamura}}, \bibinfo
  {author} {\bibfnamefont {Y.}~\bibnamefont {Ando}}, \bibinfo {author}
  {\bibfnamefont {T.}~\bibnamefont {Shinjo}}, \ and\ \bibinfo {author}
  {\bibfnamefont {M.}~\bibnamefont {Shiraishi}},\ }\href {\doibase
  10.1038/s41467-018-05611-9} {\bibfield  {journal} {\bibinfo  {journal} {Nat.
  Commun.}\ }\textbf {\bibinfo {volume} {9}},\ \bibinfo {pages} {3118}
  (\bibinfo {year} {2018})}\BibitemShut {NoStop}%
\bibitem [{\citenamefont {Chen}\ \emph
  {et~al.}(2018{\natexlab{b}})\citenamefont {Chen}, \citenamefont {Gmitra},
  \citenamefont {Vogel}, \citenamefont {Islinger}, \citenamefont {Kronseder},
  \citenamefont {Schuh}, \citenamefont {Bougeard}, \citenamefont {Fabian},
  \citenamefont {Weiss},\ and\ \citenamefont {Back}}]{Chen18_FeGaAs}%
  \BibitemOpen
  \bibfield  {author} {\bibinfo {author} {\bibfnamefont {L.}~\bibnamefont
  {Chen}}, \bibinfo {author} {\bibfnamefont {M.}~\bibnamefont {Gmitra}},
  \bibinfo {author} {\bibfnamefont {M.}~\bibnamefont {Vogel}}, \bibinfo
  {author} {\bibfnamefont {R.}~\bibnamefont {Islinger}}, \bibinfo {author}
  {\bibfnamefont {M.}~\bibnamefont {Kronseder}}, \bibinfo {author}
  {\bibfnamefont {D.}~\bibnamefont {Schuh}}, \bibinfo {author} {\bibfnamefont
  {D.}~\bibnamefont {Bougeard}}, \bibinfo {author} {\bibfnamefont
  {J.}~\bibnamefont {Fabian}}, \bibinfo {author} {\bibfnamefont
  {D.}~\bibnamefont {Weiss}}, \ and\ \bibinfo {author} {\bibfnamefont {C.~H.}\
  \bibnamefont {Back}},\ }\href {\doibase 10.1038/s41928-018-0085-1} {\bibfield
   {journal} {\bibinfo  {journal} {Nat. Electron.}\ }\textbf {\bibinfo {volume}
  {1}},\ \bibinfo {pages} {350} (\bibinfo {year}
  {2018}{\natexlab{b}})}\BibitemShut {NoStop}%
\bibitem [{\citenamefont {Emori}\ \emph {et~al.}(2014)\citenamefont {Emori},
  \citenamefont {Bauer}, \citenamefont {Woo},\ and\ \citenamefont
  {Beach}}]{Emori14}%
  \BibitemOpen
  \bibfield  {author} {\bibinfo {author} {\bibfnamefont {S.}~\bibnamefont
  {Emori}}, \bibinfo {author} {\bibfnamefont {U.}~\bibnamefont {Bauer}},
  \bibinfo {author} {\bibfnamefont {S.}~\bibnamefont {Woo}}, \ and\ \bibinfo
  {author} {\bibfnamefont {G.~S.~D.}\ \bibnamefont {Beach}},\ }\href {\doibase
  10.1063/1.4903041} {\bibfield  {journal} {\bibinfo  {journal} {Appl. Phys.
  Lett.}\ }\textbf {\bibinfo {volume} {105}},\ \bibinfo {pages} {222401}
  (\bibinfo {year} {2014})}\BibitemShut {NoStop}%
\bibitem [{\citenamefont {Mishra}\ \emph {et~al.}(2019)\citenamefont {Mishra},
  \citenamefont {Mahfouzi}, \citenamefont {Kumar}, \citenamefont {Cai},
  \citenamefont {Chen}, \citenamefont {Qiu}, \citenamefont {Kioussis},\ and\
  \citenamefont {Yang}}]{Mishra19}%
  \BibitemOpen
  \bibfield  {author} {\bibinfo {author} {\bibfnamefont {R.}~\bibnamefont
  {Mishra}}, \bibinfo {author} {\bibfnamefont {F.}~\bibnamefont {Mahfouzi}},
  \bibinfo {author} {\bibfnamefont {D.}~\bibnamefont {Kumar}}, \bibinfo
  {author} {\bibfnamefont {K.}~\bibnamefont {Cai}}, \bibinfo {author}
  {\bibfnamefont {M.}~\bibnamefont {Chen}}, \bibinfo {author} {\bibfnamefont
  {X.}~\bibnamefont {Qiu}}, \bibinfo {author} {\bibfnamefont {N.}~\bibnamefont
  {Kioussis}}, \ and\ \bibinfo {author} {\bibfnamefont {H.}~\bibnamefont
  {Yang}},\ }\href {\doibase 10.1038/s41467-018-08274-8} {\bibfield  {journal}
  {\bibinfo  {journal} {Nat. Commun.}\ }\textbf {\bibinfo {volume} {10}},\
  \bibinfo {pages} {248} (\bibinfo {year} {2019})}\BibitemShut {NoStop}%
\bibitem [{\citenamefont {Liu}\ \emph {et~al.}(2014)\citenamefont {Liu},
  \citenamefont {Lim},\ and\ \citenamefont {Urazhdin}}]{Liu14_Efield_SOT}%
  \BibitemOpen
  \bibfield  {author} {\bibinfo {author} {\bibfnamefont {R.~H.}\ \bibnamefont
  {Liu}}, \bibinfo {author} {\bibfnamefont {W.~L.}\ \bibnamefont {Lim}}, \ and\
  \bibinfo {author} {\bibfnamefont {S.}~\bibnamefont {Urazhdin}},\ }\href
  {\doibase 10.1103/PhysRevB.89.220409} {\bibfield  {journal} {\bibinfo
  {journal} {Phys. Rev. B}\ }\textbf {\bibinfo {volume} {89}},\ \bibinfo
  {pages} {220409} (\bibinfo {year} {2014})}\BibitemShut {NoStop}%
\bibitem [{\citenamefont {Sun}\ and\ \citenamefont {Xie}(2005)}]{Sun05}%
  \BibitemOpen
  \bibfield  {author} {\bibinfo {author} {\bibfnamefont {Q.-f.}\ \bibnamefont
  {Sun}}\ and\ \bibinfo {author} {\bibfnamefont {X.~C.}\ \bibnamefont {Xie}},\
  }\href {\doibase 10.1103/PhysRevB.72.245305} {\bibfield  {journal} {\bibinfo
  {journal} {Phys. Rev. B}\ }\textbf {\bibinfo {volume} {72}},\ \bibinfo
  {pages} {245305} (\bibinfo {year} {2005})}\BibitemShut {NoStop}%
\bibitem [{\citenamefont {Jin}\ \emph {et~al.}(2006)\citenamefont {Jin},
  \citenamefont {Li},\ and\ \citenamefont {Zhang}}]{Jin06}%
  \BibitemOpen
  \bibfield  {author} {\bibinfo {author} {\bibfnamefont {P.-Q.}\ \bibnamefont
  {Jin}}, \bibinfo {author} {\bibfnamefont {Y.-Q.}\ \bibnamefont {Li}}, \ and\
  \bibinfo {author} {\bibfnamefont {F.-C.}\ \bibnamefont {Zhang}},\ }\href
  {\doibase 10.1088/0305-4470/39/22/022} {\bibfield  {journal} {\bibinfo
  {journal} {J. Phys. A: Math. Gen.}\ }\textbf {\bibinfo {volume} {39}},\
  \bibinfo {pages} {7115} (\bibinfo {year} {2006})}\BibitemShut {NoStop}%
\bibitem [{\citenamefont {Shi}\ \emph {et~al.}(2006)\citenamefont {Shi},
  \citenamefont {Zhang}, \citenamefont {Xiao},\ and\ \citenamefont
  {Niu}}]{Shi06}%
  \BibitemOpen
  \bibfield  {author} {\bibinfo {author} {\bibfnamefont {J.}~\bibnamefont
  {Shi}}, \bibinfo {author} {\bibfnamefont {P.}~\bibnamefont {Zhang}}, \bibinfo
  {author} {\bibfnamefont {D.}~\bibnamefont {Xiao}}, \ and\ \bibinfo {author}
  {\bibfnamefont {Q.}~\bibnamefont {Niu}},\ }\href {\doibase
  10.1103/PhysRevLett.96.076604} {\bibfield  {journal} {\bibinfo  {journal}
  {Phys. Rev. Lett.}\ }\textbf {\bibinfo {volume} {96}},\ \bibinfo {pages}
  {076604} (\bibinfo {year} {2006})}\BibitemShut {NoStop}%
\bibitem [{\citenamefont {Sun}\ \emph {et~al.}(2008)\citenamefont {Sun},
  \citenamefont {Xie},\ and\ \citenamefont {Wang}}]{Sun08}%
  \BibitemOpen
  \bibfield  {author} {\bibinfo {author} {\bibfnamefont {Q.-f.}\ \bibnamefont
  {Sun}}, \bibinfo {author} {\bibfnamefont {X.~C.}\ \bibnamefont {Xie}}, \ and\
  \bibinfo {author} {\bibfnamefont {J.}~\bibnamefont {Wang}},\ }\href {\doibase
  10.1103/PhysRevB.77.035327} {\bibfield  {journal} {\bibinfo  {journal} {Phys.
  Rev. B}\ }\textbf {\bibinfo {volume} {77}},\ \bibinfo {pages} {035327}
  (\bibinfo {year} {2008})}\BibitemShut {NoStop}%
\end{thebibliography}%

\end{document}